

\input amssym.tex 

\def\unredoffs{}
\tolerance=1000\hfuzz=2pt
\catcode`\@=11 
\ifx\hyperdef\UNd@FiNeD\def\hyperdef#1#2#3#4{#4}\def\hyperref#1#2#3#4{#4}\def\href#1#2{#2}\fi
\magnification=1200\unredoffs\baselineskip=16pt plus 2pt minus 1pt
\def\Date#1{\vfill\leftline{#1}\tenpoint\supereject%
\footline={\hss\tenrm\hyperdef\hypernoname{page}\folio\folio\hss}}%

{\count255=\time\divide\count255 by 60 \xdef\hourmin{\number\count255}
 \multiply\count255 by-60\advance\count255 by\time
 \xdef\hourmin{\hourmin:\ifnum\count255<10 0\fi\the\count255}
}
\def\date{\number\day.\number\month.\number\year\ at \hourmin}


\def\nolabels{\def\wrlabeL##1{}\def\eqlabeL##1{}\def\reflabeL##1{}}
\def\writelabels{\def\wrlabeL##1{\leavevmode\vadjust{\rlap{\smash%
{\line{{\escapechar=` \hfill\rlap{\sevenrm\hskip.03in\string##1}}}}}}}%
\def\eqlabeL##1{{\escapechar-1\rlap{\sevenrm\hskip.05in\string##1}}}%
\def\reflabeL##1{\noexpand\llap{\noexpand\sevenrm\string\string\string##1}}}
\nolabels

\global\newcount\secno \global\secno=0
\global\newcount\meqno \global\meqno=1
\def\s@csym{}

\def\newsec#1\par{\global\advance\secno by1%
{\toks0{#1}\message{(\the\secno. \the\toks0)}}%
\global\subsecno=0\eqnres@t\let\s@csym\secsym\xdef\secn@m{\the\secno}\noindent
{\bf\hyperdef\hypernoname{section}{\the\secno}{\the\secno.} #1}%
\writetoca{{\string\hyperref{}{section}{\the\secno}{\bf \the\secno\quad}} {\bf #1}}\par%
\nobreak\medskip\nobreak\noindent\ignorespaces}
\def\eqnres@t{\xdef\secsym{\the\secno.}\global\meqno=1\bigbreak\bigskip}
\def\sequentialequations{\def\eqnres@t{\bigbreak}}\xdef\secsym{}

\global\newcount\subsecno \global\subsecno=0
\def\subsec#1\par{\global\advance\subsecno by1%
{\toks0{#1}\message{(\s@csym\the\subsecno. \the\toks0)}}%
\global\subsubsecno=0%
\ifnum\lastpenalty>9000\else\bigbreak\fi
\noindent{\it\hyperdef\hypernoname{subsection}{\secn@m.\the\subsecno}%
{\secn@m.\the\subsecno.} #1}\writetoca{\string\hskip1.45cm
{\string\hyperref{}{subsection}{\secn@m.\the\subsecno}{\secn@m.\the\subsecno.}}
{#1}}\par\nobreak\medskip\nobreak\noindent\ignorespaces}

\global\newcount\subsubsecno \global\subsubsecno=0
\def\subsubsec#1\par{\global\advance\subsubsecno by1%
{\toks0{#1}\message{(\secn@m.\the\subsecno.\the\subsubsecno. \the\toks0)}}%
\global\subsubsubsecno=0%
\ifnum\lastpenalty>9000\else\bigbreak\fi
\noindent{\it\hyperdef\hypernoname{subsubsection}{\secn@m.\the\subsecno\the\subsubsecno}%
{\secn@m.\the\subsecno.\the\subsubsecno.} #1}
\par\nobreak\medskip\nobreak\noindent\ignorespaces}

\global\newcount\subsubsubsecno \global\subsubsubsecno=0
\def\subsubsubsec#1\par{\global\advance\subsubsubsecno by1%
{\toks0{#1}\message{(\secn@m.\the\subsecno.\the\subsubsecno.\the\subsubsubsecno \the\toks0)}}%
\ifnum\lastpenalty>9000\else\bigbreak\fi
\noindent{\it\hyperdef\hypernoname{subsubsection}{\secn@m.\the\subsecno\the\subsubsecno\the\subsubsubsecno}%
{\secn@m.\the\subsecno.\the\subsubsecno.\the\subsubsubsecno.} #1}%
\par\nobreak\medskip\nobreak\noindent\ignorespaces}


\def\newnewsec#1#2\par{\global\advance\secno by1%
{\toks0{#2}\message{(\secn@m. \the\toks0)}}%
\global\subsecno=0\global\subsubsecno=0\eqnres@t\let\s@csym\secsym\xdef\secn@m{\the\secno}\noindent
\ifnum\lastpenalty>9000\else\bigbreak\fi
\noindent{\bf\hyperdef\hypernoname{section}{\secn@m}{\secn@m.} #2}%
\writetoca{{\string\hyperref{}{section}{\the\secno}{\bf \the\secno\quad}} {\bf #2}}
\DefWarn#1%
\xdef#1{\noexpand\hyperref{}{section}{\the\secno}%
{\the\secno}}\writedef{#1\leftbracket#1}\wrlabeL{#1=#1}%
\par\nobreak\medskip\nobreak\noindent\ignorespaces}

\def\newsubsec#1#2\par{\global\advance\subsecno by1%
{\toks0{#2}\message{(\secn@m.\the\subsecno. \the\toks0)}}%
\global\subsubsecno=0%
\ifnum\lastpenalty>9000\else\bigbreak\fi
\noindent{\it\hyperdef\hypernoname{subsection}{\secn@m.\the\subsecno}%
{\secn@m.\the\subsecno.} #2}
\DefWarn#1%
\xdef#1{\noexpand\hyperref{}{subsection}{\secn@m.\the\subsecno}%
{\secn@m.\the\subsecno}}\writedef{#1\leftbracket#1}\wrlabeL{#1=#1}%
\writetoca{\string\hskip1.45cm
{\string\hyperref{}{subsection}{\secn@m.\the\subsecno}{\secn@m.\the\subsecno.}}
{#2}}%
\par\nobreak\medskip\nobreak\noindent\ignorespaces}

\def\newsubsecstar#1#2\par{\global\advance\subsecno by1%
{\toks0{#2}\message{(\secn@m.\the\subsecno. \the\toks0)}}%
\global\subsubsecno=0%
\ifnum\lastpenalty>9000\else\bigbreak\fi
\noindent{\it\hyperdef\hypernoname{subsection}{\secn@m.\the\subsecno}%
{\secn@m.\the\subsecno.} #2}
\DefWarn#1%
\xdef#1{\noexpand\hyperref{}{subsection}{\secn@m.\the\subsecno}%
{\secn@m.\the\subsecno}}\writedef{#1\leftbracket#1}\wrlabeL{#1=#1}%
\par\nobreak\medskip\nobreak\noindent\ignorespaces}

\def\newsubsubsec#1#2\par{\global\advance\subsubsecno by1%
{\toks0{#2}\message{(\secn@m.\the\subsecno.\the\subsubsecno. \the\toks0)}}%
\global\subsubsubsecno=0%
\ifnum\lastpenalty>9000\else\bigbreak\fi
\noindent{\it\hyperdef\hypernoname{subsubsection}{\secn@m.\the\subsecno.\the\subsubsecno}%
{\secn@m.\the\subsecno.\the\subsubsecno.} #2}
\DefWarn#1%
\xdef#1{\noexpand\hyperref{}{subsubsection}{\secn@m.\the\subsecno.\the\subsubsecno}%
{\secn@m.\the\subsecno.\the\subsubsecno}}\writedef{#1\leftbracket#1}\wrlabeL{#1=#1}%
\par\nobreak\medskip\nobreak\noindent\ignorespaces}

\def\newsubsubsubsec#1#2\par{\global\advance\subsubsubsecno by1%
{\toks0{#2}\message{(\secn@m.\the\subsecno.\the\subsubsecno.\the\subsubsubsecno \the\toks0)}}%
\ifnum\lastpenalty>9000\else\bigbreak\fi
\noindent{\it\hyperdef\hypernoname{subsubsection}{\secn@m.\the\subsecno\the\subsubsecno\the\subsubsubsecno}%
{\secn@m.\the\subsecno.\the\subsubsecno.\the\subsubsubsecno.} #2}
\DefWarn#1%
\xdef#1{\noexpand\hyperref{}{subsubsubsection}{\secn@m.\the\subsecno.\the\subsubsecno.\the\subsubsubsecno}%
{\secn@m.\the\subsecno.\the\subsubsecno.\the\subsubsubsecno}}\writedef{#1\leftbracket#1}\wrlabeL{#1=#1}%
\par\nobreak\medskip\nobreak\noindent\ignorespaces}

\def\appendix#1#2{\global\meqno=1\global\subsecno=0\global\subsubsecno=0\xdef\secsym{\hbox{#1.}}%
\bigbreak\bigskip\noindent{\bf Appendix \hyperdef\hypernoname{appendix}{#1}%
{#1.} #2}{\toks0{(#1. #2)}\message{\the\toks0}}%
\xdef\s@csym{#1.}\xdef\secn@m{#1}%
\writetoca{{\string\hyperref{}{appendix}{#1}{\bf {#1}\quad}} {\bf #2}}%
\par\nobreak\medskip\nobreak}

%
\def\checkm@de#1#2{\ifmmode{\def\f@rst##1{##1}\hyperdef\hypernoname{equation}%
{#1}{#2}}\else\hyperref{}{equation}{#1}{#2}\fi}
\def\eqnn#1{\DefWarn#1\xdef #1{(\noexpand\relax\noexpand\checkm@de%
{\s@csym\the\meqno}{\secsym\the\meqno})}%
\wrlabeL#1\writedef{#1\leftbracket#1}\global\advance\meqno by1}
\def\f@rst#1{\c@t#1a\em@ark}\def\c@t#1#2\em@ark{#1}
\def\eqna#1{\DefWarn#1\wrlabeL{#1$\{\}$}%
\xdef #1##1{(\noexpand\relax\noexpand\checkm@de%
{\s@csym\the\meqno\noexpand\f@rst{##1}1}{\hbox{$\secsym\the\meqno##1$}})}
\writedef{#1\numbersign1\leftbracket#1{\numbersign1}}\global\advance\meqno by1}
\def\eqn#1#2{\DefWarn#1%
\xdef #1{(\noexpand\hyperref{}{equation}{\s@csym\the\meqno}%
{\secsym\the\meqno})}$$#2\eqno(\hyperdef\hypernoname{equation}%
{\s@csym\the\meqno}{\secsym\the\meqno})\eqlabeL#1$$%
\writedef{#1\leftbracket#1}\global\advance\meqno by1}
\def\xeqn{\expandafter\xe@n}\def\xe@n(#1){#1}
\def\xeqna#1{\expandafter\xe@n#1}
\def\eqns#1{(\e@ns #1{\hbox{}})}
\def\e@ns#1{\ifx\UNd@FiNeD#1\message{eqnlabel \string#1 is undefined.}%
\xdef#1{(?.?)}\fi{\let\hyperref=\relax\xdef\next{#1}}%
\ifx\next\em@rk\def\next{}\else%
\ifx\next#1\xeqn#1\else\def\n@xt{#1}\ifx\n@xt\next#1\else\xeqna#1\fi
\fi\let\next=\e@ns\fi\next}
\def\DefWarn#1{}
%
\newskip\footskip\footskip14pt plus 1pt minus 1pt 
\def\footnotefont{\ninepoint}\def\f@t#1{\footnotefont #1\@foot}
\def\f@@t{\baselineskip\footskip\bgroup\footnotefont\aftergroup\@foot\let\next}
\setbox\strutbox=\hbox{\vrule height9.5pt depth4.5pt width0pt}
\global\newcount\ftno \global\ftno=0
\def\foot{\global\advance\ftno by1\def\foot@rg{\hyperref{}{footnote}%
{\the\ftno}{\the\ftno}\xdef\foot@rg{\noexpand\hyperdef\noexpand\hypernoname%
{footnote}{\the\ftno}{\the\ftno}}}\footnote{$^{\foot@rg}$}}
%
%
%
\global\newcount\refno \global\refno=1
\newwrite\rfile
\def\ref{[\hyperref{}{reference}{\the\refno}{\the\refno}]\nref}
\def\nref#1{\DefWarn#1%
\xdef#1{[\noexpand\hyperref{}{reference}{\the\refno}{\the\refno}]}%
\writedef{#1\leftbracket#1}%
\ifnum\refno=1\immediate\openout\rfile=\jobname.refs\fi
\chardef\wfile=\rfile\immediate\write\rfile{\noexpand\item{[\noexpand\hyperdef%
\noexpand\hypernoname{reference}{\the\refno}{\the\refno}]\ }%
\reflabeL{#1\hskip.31in}\pctsign}\global\advance\refno by1\findarg}
\def\findarg#1#{\begingroup\obeylines\newlinechar=`\^^M\pass@rg}
{\obeylines\gdef\pass@rg#1{\writ@line\relax #1^^M\hbox{}^^M}%
\gdef\writ@line#1^^M{\expandafter\toks0\expandafter{\striprel@x #1}%
\edef\next{\the\toks0}\ifx\next\em@rk\let\next=\endgroup\else\ifx\next\empty%
\else\immediate\write\wfile{\the\toks0}\fi\let\next=\writ@line\fi\next\relax}}
\def\striprel@x#1{} \def\em@rk{\hbox{}}
\def\lref{\begingroup\obeylines\lr@f}
\def\lr@f#1#2{\DefWarn#1\gdef#1{\let#1=\UNd@FiNeD\ref#1{#2}}\endgroup\unskip}
\def\semi{;\hfil\break}
\def\addref#1{\immediate\write\rfile{\noexpand\item{}#1}} 
\def\listrefs{\vfill\supereject\immediate\closeout\rfile\writestoppt
\baselineskip=\footskip\centerline{{\bf References}}\bigskip{\parindent=20pt%
\frenchspacing\escapechar=` \input \jobname.refs\vfill\eject}\nonfrenchspacing}
\def\startrefs#1{\immediate\openout\rfile=\jobname.refs\refno=#1}
\def\xref{\expandafter\xr@f}\def\xr@f[#1]{#1}
\def\refs#1{\count255=1[\r@fs #1{\hbox{}}]}
\def\r@fs#1{\ifx\UNd@FiNeD#1\message{reflabel \string#1 is undefined.}%
\nref#1{need to supply reference \string#1.}\fi%
\vphantom{\hphantom{#1}}{\let\hyperref=\relax\xdef\next{#1}}%
\ifx\next\em@rk\def\next{}%
\else\ifx\next#1\ifodd\count255\relax\xref#1\count255=0\fi%
\else#1\count255=1\fi\let\next=\r@fs\fi\next}
%

%
\newwrite\ffile\global\newcount\figno \global\figno=1
\def\fig{fig.~\hyperref{}{figure}{\the\figno}{\the\figno}\nfig}
\def\nfig#1{\DefWarn#1%
\xdef#1{fig.~\noexpand\hyperref{}{figure}{\the\figno}{\the\figno}}%
\writedef{#1\leftbracket fig.\noexpand~\xfig#1}%
\ifnum\figno=1\immediate\openout\ffile=\jobname.figs\fi\chardef\wfile=\ffile%
{\let\hyperref=\relax
\immediate\write\ffile{\noexpand\medskip\noexpand\item{Fig.\ %
\noexpand\hyperdef\noexpand\hypernoname{figure}{\the\figno}{\the\figno}. }
\reflabeL{#1\hskip.55in}\pctsign}}\global\advance\figno by1\findarg}
\def\xfig{\expandafter\xf@g}\def\xf@g fig.\penalty\@M\ {}
\def\figs#1{figs.~\f@gs #1{\hbox{}}}
\def\f@gs#1{{\let\hyperref=\relax\xdef\next{#1}}\ifx\next\em@rk\def\next{}\else
\ifx\next#1\xfig #1\else#1\fi\let\next=\f@gs\fi\next}
%
\def\figin{\epsfcheck\figin}\def\figins{\epsfcheck\figins}
\def\epsfcheck{\ifx\epsfbox\UnDeFiNeD
\message{(NO epsf.tex, FIGURES WILL BE IGNORED)}
\gdef\figin##1{\vskip2in}\gdef\figins##1{\hskip.5in}
\else\message{(FIGURES WILL BE INCLUDED)}%
\gdef\figin##1{##1}\gdef\figins##1{##1}\fi}
\def\figinsert{\goodbreak\topinsert}
\def\ifig#1#2#3{\DefWarn#1\xdef#1{fig.~\the\figno}
\writedef{#1\leftbracket fig.\noexpand~\the\figno}%
\figinsert\figin{\centerline{#3}}
\smallskip
\leftskip=0pt \rightskip=0pt
\baselineskip12pt\noindent
{{\bf Fig.~\the\figno}\ \ninepoint #2}
\medskip
\global\advance\figno by1\par\endinsert}
\newwrite\lfile
{\escapechar-1\xdef\pctsign{\string\%}\xdef\leftbracket{\string\{}
\xdef\rightbracket{\string\}}\xdef\numbersign{\string\#}}
\def\writedefs{\immediate\openout\lfile=label.defs \def\writedef##1{%
{\let\hyperref=\relax\let\hyperdef=\relax\let\hypernoname=\relax
 \immediate\write\lfile{\string\checkdef\string##1\rightbracket}}}}%
\def\writestop{\def\writestoppt{\immediate\write\lfile{\string\pageno
 \the\pageno\string\startrefs\leftbracket\the\refno\rightbracket
 \string\def\string\secsym\leftbracket\secsym\rightbracket
 \string\secno\the\secno\string\meqno\the\meqno}\immediate\closeout\lfile}}
\def\writestoppt{}\def\writedef#1{}

\def\seclab#1\par{\DefWarn#1%
\xdef #1{\noexpand\hyperref{}{section}{\the\secno}{\the\secno}}%
\writedef{#1\leftbracket#1}\wrlabeL{#1=#1}\par%
\nobreak\medskip\nobreak\noindent\ignorespaces}
\def\subseclab#1\par{\DefWarn#1%
\xdef #1{\noexpand\hyperref{}{subsection}{\the\secno.\the\subsecno}%
{\the\secno.\the\subsecno}}\writedef{#1\leftbracket#1}\wrlabeL{#1=#1}\par%
\nobreak\medskip\nobreak\noindent\ignorespaces}
\def\subsubseclab#1\par{\DefWarn#1%
\xdef#1{\noexpand\hyperref{}{subsubsection}{\the\secno.\the\subsecno.\the\subsubsecno}%
{\the\secno.\the\subsecno.\the\subsubsecno}}\writedef{#1\leftbracket#1}\wrlabeL{#1=#1}\par%
\nobreak\medskip\nobreak\noindent\ignorespaces}
\def\applab#1\par{\DefWarn#1%
\xdef#1{\noexpand\hyperref{}{appendix}{\secn@m}{\secn@m}}%
\writedef{#1\leftbracket#1}\wrlabeL{#1=#1}%
\par\nobreak\medskip\nobreak\noindent\ignorespaces}
\def\appsublab#1{\DefWarn#1%
\xdef #1{\noexpand\hyperref{}{appendix}{\secn@m.\the\subsecno}{\secn@m.\the\subsecno}}%
\writedef{#1\leftbracket#1}\wrlabeL{#1=#1}}
\newwrite\tfile \def\writetoca#1{}
\def\leaderfill{\leaders\hbox to 1em{\hss.\hss}\hfill}
\def\writetoc{\immediate\openout\tfile=\jobname.toc
   \def\writetoca##1{{\edef\next{\write\tfile{\noindent ##1
   \string\leaderfill{
   \string\hyperref{}{page}{\noexpand\number\pageno}%
   {\noexpand\number\pageno}} \par}}\next}}
}
\newread\ch@ckfile
\def\listtoc{\immediate\closeout\tfile\immediate\openin\ch@ckfile=\jobname.toc
\ifeof\ch@ckfile\message{no file \jobname.toc, no table of contents this pass}%
\else\closein\ch@ckfile\centerline{\bf Contents}\nobreak\medskip%
{\baselineskip=15.5pt\footnotefont\parskip=0pt\catcode`\@=11\input\jobname.toc
\catcode`\@=12\bigbreak\bigskip}\fi}
\catcode`\@=12 
\def\tenpoint{\def\rm{\fam0\tenrm}
\textfont0=\tenrm \scriptfont0=\sevenrm \scriptscriptfont0=\fiverm
\textfont1=\teni  \scriptfont1=\seveni  \scriptscriptfont1=\fivei
\textfont2=\tensy \scriptfont2=\sevensy \scriptscriptfont2=\fivesy
\textfont\itfam=\tenit \def\it{\fam\itfam\tenit}\def\footnotefont{\ninepoint}%
\textfont\bffam=\tenbf \def\bf{\fam\bffam\tenbf}\def\sl{\fam\slfam\tensl}\rm}
\font\ninerm=cmr9 \font\sixrm=cmr6 \font\ninei=cmmi9 \font\sixi=cmmi6
\font\ninesy=cmsy9 \font\sixsy=cmsy6 \font\ninebf=cmbx9
\font\nineit=cmti9 \font\ninesl=cmsl9 \skewchar\ninei='177
\skewchar\sixi='177 \skewchar\ninesy='60 \skewchar\sixsy='60
\def\ninepoint{\def\rm{\fam0\ninerm}
\textfont0=\ninerm \scriptfont0=\sixrm \scriptscriptfont0=\fiverm
\textfont1=\ninei \scriptfont1=\sixi \scriptscriptfont1=\fivei
\textfont2=\ninesy \scriptfont2=\sixsy \scriptscriptfont2=\fivesy
\textfont\itfam=\ninei \def\it{\fam\itfam\nineit}\def\sl{\fam\slfam\ninesl}%
\textfont\bffam=\ninebf \def\bf{\fam\bffam\ninebf}\rm}
%
\hyphenation{anom-aly anom-alies coun-ter-term coun-ter-terms}

\def\tikzcaption#1#2{\DefWarn#1\xdef#1{Fig.~\the\figno}
\writedef{#1\leftbracket Fig.\noexpand~\the\figno}%
{
\smallskip
\leftskip=20pt \rightskip=20pt \baselineskip12pt\noindent
{{\bf Fig.~\the\figno}\ \ninepoint #2}
\bigskip
\global\advance\figno by1 \par}}

\def\ntoalpha#1{%
\ifcase#1%
@%
\or A\or B\or C\or D\or E\or F\or G\or H\or I\or J\or K\or L\or M%
\fi
}

\global\newcount\appno \global\appno=1
\def\applab#1{\xdef #1{\ntoalpha{\appno}}\writedef{#1\leftbracket#1}\wrlabeL{#1=#1}
\global\advance\appno by1}

\def\preprint#1 #2\par{\rightline{\vbox{\baselineskip12pt\hbox{#1}\hbox{#2}}}\vskip2cm}
%
\def\title#1\par{\centerline{\bf #1}\nopagenumbers\pageno=0}
\def\author#1\par{\bigskip\bigskip\centerline{#1}}

\newcount\addressno

\def\email#1#2{
\footnote{\null}{\kern-\parindent \llap{$^#1$\hskip1pt}email: #2}}

\def\startcenter{%
  \par
  \begingroup
  \leftskip=0pt plus 1fil
  \rightskip=\leftskip
  \parindent=0pt
  \parfillskip=0pt
}
\def\stopcenter{\endgroup}

\def\address{\bigskip%
  \ifnum\the\addressno=0\else\stopcenter\endgroup\fi
  \advance\addressno by 1%
  \begingroup
  \startcenter
  \it
  \obeylines
  \addressAux
}
\def\addressAux#1{#1}

\def\abstract{\stopcenter\endgroup\bigskip\bigskip\noindent}

\def\Dsl{\,\raise.15ex\hbox{/}\mkern-13.5mu D} 
\def\dsl{\raise.15ex\hbox{/}\kern-.57em\partial}
 \def\Tr{{\rm Tr}}
\def\boxeqn#1{\vcenter{\vbox{\hrule\hbox{\vrule\kern3pt\vbox{\kern3pt
	\hbox{${\displaystyle #1}$}\kern3pt}\kern3pt\vrule}\hrule}}}


\def\ap{{\alpha^{\prime}}}

\def\g{{\gamma}}
\def\d{{\delta}}

\def\l{\lambda}

\def\s{{\sigma}}
\def\t{{\theta}}

\def\half{{1\over 2}}

\def\({\left(}
\def\){\right)}

\def\cI{{\cal I}}

\def\bV{{\Bbb V}}

\def\bW{{\Bbb W}}
\def\bF{{\Bbb F}}

\def\AYM{A^{\rm SYM}}


\def\len#1{{%
\def\Dlen{\left|\mkern-1mu #1\mkern -0.5mu\right|}%
\def\Sslen{\left|\mkern-1.3mu #1\mkern -1.3mu\right|}%
\def\SSlen{\left|\mkern-2.8mu #1\mkern-1.3mu\right|}%
\mathchoice{\Dlen}{\Dlen}{\Sslen}{\SSlen}}}

\def\sfrac#1/#2{\kern.1em\raise.5ex\hbox{\the\scriptfont0 #1}%
\kern-.1em/\kern-.15em\lower.25ex\hbox{\the\scriptfont0 #2}}

\font\tenshuffle=shuffle10 \font\sevenshuffle=shuffle7 \font\fiveshuffle=shuffle7 at 5pt
\def\shuffle{{%
\def\Dshuffle{\mathbin{\hbox{\tenshuffle\char'001}}}%
\def\Sshuffle{\mathbin{\hbox{\sevenshuffle\char'001}}}%
\def\SSshuffle{\mathbin{\hbox{\fiveshuffle\char'001}}}%
\mathchoice{\Dshuffle}{\Dshuffle}{\Sshuffle}{\SSshuffle}}}


\def\qed{\hbox{\hskip 3pt
\vbox{\hrule\hbox to 7pt{\vrule height 7pt\hfill\vrule}
\hrule}}\hskip3pt}

\overfullrule=0pt\relax

\frenchspacing

\def\checkdef#1#2{
\ifx\UndeFined#1%
	\def#1{#2}
\else
	\immediate\write16{*** BUG ***: the label \string#1 is already defined ***}
\fi
}
\newread\instream

\openin\instream= label.defs
\ifeof\instream\message{No labels in advance yet. Wait till next pass.}
\else\closein\instream \input label.defs
\fi
\writedefs

\def\arXiv:#1].{\hepthStrip#1 \nil}
\def\hepthStrip#1 #2\nil{\href{http://arxiv.org/abs/#1}{arXiv:#1 #2\unskip}].}


\def\Tr{{\rm Tr}}
\def\AYM{A^{\rm SYM}}
\def\AFq{{A_{\z_2}}}
\def\Astring{A^{\rm string}}
\def\stirling{\atopwithdelims[]}
\def\si{\sigma}
\def\z{f}
\def\textbf#1{{\bf #1}}
\def\Wordd(#1,#2){W_{#1#2}}
\def\Word(#1,#2,#3){W_{#1#2#3}}
\def\Wordq(#1,#2,#3,#4){W_{#1#2#3#4}}
\def\AFqf(#1,#2,#3,#4,#5){A^{\z_2}_{#1#2#3#4#5}}
\def\Du(#1){{D_{\{#1\}}}}
\def\cE{\mathord{\cal E}}
\def\stirling{\atopwithdelims[]}
\def\QED{\hfill\qed}
\def\ss(#1,#2){s_{#1#2}}

\title KK-like relations of $\ap$ corrections to disk amplitudes

\author
Carlos R. Mafra\email{\dagger}{c.r.mafra@soton.ac.uk}

\address
Mathematical Sciences and STAG Research Centre, University of Southampton,
Highfield, Southampton, SO17 1BJ, UK

\abstract
Inspired by the definition of color-dressed amplitudes in string theory, we define
analogous {\it color-dressed permutations} replacing the color-ordered string amplitudes by
their corresponding permutations. Decomposing the color traces into symmetrized traces and
structure constants, the color-dressed permutations
define {\it BRST-invariant permutations}, which we show
are elements of the inverse Solomon descent algebra and we find a closed formula for them.
We then present evidence that these permutations encode KK-like relations among the
different $\ap$ corrections to disk amplitudes refined by their motivic MZV content.
In particular, the number of linearly independent amplitudes at a given $\ap$ order and motivic MZV content
is given by (sums of) Stirling cycle numbers.
In addition, we show how the superfield expansion of BRST invariants
of the pure spinor formalism corresponding to $\ap^2\z_2$ corrections
is encoded in the descent algebra.

\Date{August 2021}

\lref\schwarzrev{
	J.H.~Schwarz,
	``Superstring Theory,''
	Phys. Rept. \textbf{89}, 223-322 (1982)
}

\lref\schabinger{
R.M.~Schabinger,
``One-Loop N = 4 Super Yang-Mills Scattering Amplitudes to All Orders in the Dimensional Regularization Parameter,''
[arXiv:1103.2769 [hep-th]].
}

\lref\BGapwww{
{\tt http://repo.or.cz/BGap.git}
}

\lref\brownmot{
	F.~Brown, ``On the decomposition of motivic multiple zeta values.''
	In Galois–Teichmüller theory and arithmetic geometry, pp. 31-58. Mathematical Society of Japan, 2012
	[arXiv:1102.1310 [math.NT]]
}

\lref\greenteight{
	M.~B.~Green and J.~H.~Schwarz,
	``Supersymmetrical Dual String Theory. 2. Vertices and Trees,''
	Nucl. Phys. B \textbf{198}, 252-268 (1982)
}

\lref\PScomb{
	C.R.~Mafra,
	``Planar binary trees in scattering amplitudes.''
	Algebraic Combinatorics, Resurgence, Moulds and Applications (CARMA) (2020): 349-365.
	[arXiv:2011.14413 [math.CO]].
}

\lref\olpriv{
O. Schlotterer, private communication.
}

\lref\momKern{
	N.~E.~J.~Bjerrum-Bohr, P.~H.~Damgaard, T.~Sondergaard and P.~Vanhove,
	``The Momentum Kernel of Gauge and Gravity Theories,''
	JHEP {\bf 1101}, 001 (2011)=.
	[arXiv:1010.3933].
}
\lref\oldMomKer{
	Z.~Bern, L.~J.~Dixon, M.~Perelstein and J.~S.~Rozowsky,
	``Multileg one loop gravity amplitudes from gauge theory,''
	Nucl.\ Phys.\ B {\bf 546}, 423 (1999).
	[hep-th/9811140].
}
\lref\BohrMomKer{
	N.~E.~J.~Bjerrum-Bohr, P.~H.~Damgaard, B.~Feng and T.~Sondergaard,
	``Gravity and Yang-Mills Amplitude Relations,''
	Phys.\ Rev.\ D {\bf 82}, 107702 (2010).
	[arXiv:1005.4367].
}

\lref\mizeraKLT{
	S.~Mizera,
	``Inverse of the String Theory KLT Kernel,''
	JHEP \textbf{06}, 084 (2017)
	[arXiv:1610.04230].
}

\lref\Polylogs{
	J.~Broedel, O.~Schlotterer and S.~Stieberger,
	``Polylogarithms, Multiple Zeta Values and Superstring Amplitudes,''
	Fortsch.\ Phys.\  {\bf 61}, 812 (2013).
	[arXiv:1304.7267].
}
\lref\SYM{
	L.~Brink, J.~H.~Schwarz and J.~Scherk,
	``Supersymmetric Yang-Mills Theories,''
	Nucl. Phys. B \textbf{121}, 77-92 (1977)
}

\lref\semiZ{
	J.~J.~M.~Carrasco, C.R.~Mafra and O.~Schlotterer,
	``Semi-abelian Z-theory: NLSM$+\phi^{3}$ from the open string,''
	JHEP \textbf{08}, 135 (2017)
	[arXiv:1612.06446].
}

\lref\boels{
	R.~H.~Boels and R.~S.~Isermann,
  	``Yang-Mills amplitude relations at loop level from non-adjacent BCFW shifts,''
	[arXiv:1110.4462].
}

\lref\NLSM{
	J.J.M.~Carrasco, C.R.~Mafra and O.~Schlotterer,
	``Abelian Z-theory: NLSM amplitudes and $\alpha$'-corrections from the open string,''
	JHEP \textbf{06}, 093 (2017)
	[arXiv:1608.02569].
}

\lref\FORM{
	J.A.M.~Vermaseren,
  	``New features of FORM,''
	[math-ph/0010025].
\semi
	J.~Kuipers, T.~Ueda, J.A.M.~Vermaseren and J.~Vollinga,
	``FORM version 4.0,''
	Comput.\ Phys.\ Commun.\  {\bf 184}, 1453 (2013).
	[arXiv:1203.6543 [cs.SC]].
}

\lref\nptMethod{
	C.R.~Mafra, O.~Schlotterer, S.~Stieberger and D.~Tsimpis,
	``A recursive method for SYM n-point tree amplitudes,''
	Phys.\ Rev.\ D {\bf 83}, 126012 (2011).
	[arXiv:1012.3981].
}

\lref\BCJ{
	Z.~Bern, J.J.M.~Carrasco and H.~Johansson,
  	``New Relations for Gauge-Theory Amplitudes,''
	Phys.\ Rev.\ D {\bf 78}, 085011 (2008).
	[arXiv:0805.3993 [hep-ph]].
}

\lref\monodStieberger{
	S.~Stieberger,
  	``Open \& Closed vs. Pure Open String Disk Amplitudes,''
	[arXiv:0907.2211].
}
\lref\monodVanhove{
	N.E.J.~Bjerrum-Bohr, P.H.~Damgaard and P.~Vanhove,
  	``Minimal Basis for Gauge Theory Amplitudes,''
	Phys.\ Rev.\ Lett.\  {\bf 103}, 161602 (2009).
	[arXiv:0907.1425].
}
\lref\motivic{
	O.~Schlotterer and S.~Stieberger,
  	``Motivic Multiple Zeta Values and Superstring Amplitudes,''
	J.\ Phys.\ A {\bf 46}, 475401 (2013).
	[arXiv:1205.1516].
}

\lref\rugpriv{
	R. Bandiera, private communication.
}
\lref\broedeldixon{
	J.~Broedel and L.~J.~Dixon,
  	``Color-kinematics duality and double-copy construction for amplitudes from higher-dimension operators,''
	JHEP {\bf 1210}, 091 (2012).
	[arXiv:1208.0876].
}
\lref\oprisa{
	D.~Oprisa and S.~Stieberger,
  	``Six gluon open superstring disk amplitude, multiple hypergeometric series and Euler-Zagier sums,''
	[hep-th/0509042].
}
\lref\BGap{
	C.R.~Mafra and O.~Schlotterer,
  	``Non-abelian $Z$-theory: Berends-Giele recursion for the $\alpha'$-expansion of disk integrals,''
	[arXiv:1609.07078].
	{\tt http://repo.or.cz/BGap.git}
}
\lref\medina{
	R.~Medina, F.~T.~Brandt and F.~R.~Machado,
  	``The Open superstring five point amplitude revisited,''
	JHEP {\bf 0207}, 071 (2002).
	[hep-th/0208121].
\semi
	L.~A.~Barreiro and R.~Medina,
  	``5-field terms in the open superstring effective action,''
	JHEP {\bf 0503}, 055 (2005).
	[hep-th/0503182].
}

\lref\multigluon{
	S.~Stieberger and T.~R.~Taylor,
  	``Multi-Gluon Scattering in Open Superstring Theory,''
	Phys.\ Rev.\ D {\bf 74}, 126007 (2006).
	[hep-th/0609175].
}
\lref\ragoucy{
	J.~M.~Drummond and E.~Ragoucy,
  	``Superstring amplitudes and the associator,''
	JHEP {\bf 1308}, 135 (2013).
	[arXiv:1301.0794].
}

\lref\drinfeld{
	J.~Broedel, O.~Schlotterer, S.~Stieberger and T.~Terasoma,
	``All order $\alpha^{\prime}$-expansion of superstring trees from the Drinfeld associator,''
	Phys.\ Rev.\ D {\bf 89}, no. 6, 066014 (2014).
	[arXiv:1304.7304].
	{\tt http://mzv.mpp.mpg.de}
}

\lref\copenhagen{
	N.~E.~J.~Bjerrum-Bohr, P.~H.~Damgaard, H.~Johansson and T.~Sondergaard,
  	``Monodromy--like Relations for Finite Loop Amplitudes,''
	JHEP {\bf 1105}, 039 (2011).
	[arXiv:1103.6190].
}

\lref\bilal{
	A.~Bilal,
	``Higher derivative corrections to the nonAbelian Born-Infeld action,''
	Nucl.\ Phys.\ B {\bf 618}, 21 (2001).
	[hep-th/0106062].
}

\lref\KKref{
	R.~Kleiss and H.~Kuijf,
	``Multi - Gluon Cross-sections and Five Jet Production at Hadron Colliders,''
	Nucl.\ Phys.\ B {\bf 312}, 616 (1989)..
}

\lref\psf{
 	N.~Berkovits,
	``Super-Poincare covariant quantization of the superstring,''
	JHEP {\bf 0004}, 018 (2000)
	[arXiv:hep-th/0001035].
}

\lref\BerendsME{
	F.A.~Berends and W.T.~Giele,
	``Recursive Calculations for Processes with n Gluons,''
	Nucl.\ Phys.\ B {\bf 306}, 759 (1988).
}

\lref\tnn{
	D.E. Knuth, ``Two notes on notation'', Amer. Math. Monthly 99 (1992), no.\ 5, 403--422
	[math/9205211].
}
\lref\knuthconcrete{
	R. Graham, D.E. Knuth, and O. Patashnik,
 	``Concrete Mathematics: A Foundation for Computer Science'',
	Addison-Wesley Longman Publishing Co., Inc.,
	Boston, MA, USA, (1994).
}
\lref\flas{
	H.~Frost, C.R.~Mafra and L.~Mason,
	``A Lie bracket for the momentum kernel,''
	[arXiv:2012.00519].
}

\lref\malvenutoreutenauer{
C.~Malvenuto, and C.~Reutenauer,
``Duality between quasi-symmetrical functions and the solomon descent algebra'',
Journal of Algebra, 177(3), (1995) pp.967-982.
}

\lref\Ree{
	R.~Ree, ``Lie elements and an algebra associated with shuffles'',
	Ann.Math. {\bf 62}, No.2 (1958), 210--220.
}
\lref\lothaire{
	Lothaire, M., ``Combinatorics on Words'',
	(Cambridge Mathematical Library), Cambridge University Press (1997).
}

\lref\BGschocker{
	M. Schocker,
	``Lie elements and Knuth relations,'' Canad. J. Math. {\bf 56} (2004), 871-882.
	[math/0209327].
}

\lref\mackey{
	L.~Solomon, ``A Mackey formula in the group ring of a Coxeter group,''
	Journal of Algebra, 41(2), (1976) 255-264.
}

\lref\schockerSolomon{
	M.~Schocker,
	``The descent algebra of the symmetric group,''
	Representations of finite dimensional algebras and related topics in Lie theory and geometry 40 (2004): 145-161.
}

\lref\thibon{
	J.-Y. Thibon,
	``Lie idempotents in descent algebras''
	(lecture notes), Workshop on Hopf Algebras and Props, Boston, March 5 - 9, 2007 Clay Mathematics Institute.
}

\lref\oneloopbb{
	C.R.~Mafra and O.~Schlotterer,
  	``The Structure of n-Point One-Loop Open Superstring Amplitudes,''
	JHEP {\bf 1408}, 099 (2014).
	[arXiv:1203.6215].
}
\lref\polchinski{
	J.~Polchinski,
  	``String theory. Vol. 1: An introduction to the bosonic string,''
	{\it  Cambridge, UK: Univ. Pr. (1998) 402 p}
}

\lref\Vcolor{
	T.~van Ritbergen, A.~N.~Schellekens and J.~A.~M.~Vermaseren,
  	``Group theory factors for Feynman diagrams,''
	Int.\ J.\ Mod.\ Phys.\ A {\bf 14}, 41 (1999).
	[hep-ph/9802376].
}
\lref\ctrace{
	R.~Bandiera and C.R.~Mafra,
	``A closed-formula solution to the color-trace decomposition problem,''
	[arXiv:2009.02534 [math.CO]].
}
\lref\ddm{
	V.~Del Duca, L.~J.~Dixon and F.~Maltoni,
  	``New color decompositions for gauge amplitudes at tree and loop level,''
	Nucl.\ Phys.\ B {\bf 571}, 51 (2000).
	[hep-ph/9910563].
}

\lref\Reutenauer{
	C.~Reutenauer,
	``Free Lie Algebras,''
	London Mathematical Society Monographs, 1993
}
\lref\PBWReutenauer{
	C.~Reutenauer,
	``Theorem of Poincar\'e-Birkhoff-Witt, logarithm and
	symmetric group representations of degrees equal to
	Stirling numbers''. In Combinatoire \'enum\'erative (pp. 267-284) 1986. Springer, Berlin, Heidelberg.
}
\lref\schatz{
	R.~Bandiera, F.~Schaetz, ``Eulerian idempotent, pre-Lie logarithm and combinatorics of trees''.
	arXiv:1702.08907.
}
\lref\Loday{
	J.L.~Loday, ``S\'erie de Hausdorff, idempotents Eul\'eriens et algebres de Hopf''.
	Exp. Math. {\bf 12} (1994), 165-178.
}
\lref\solomon{
	L.~Solomon ``On the Poincar\'e-Birkhoff-Witt theorem''.
	Journal of Combinatorial Theory. 1968 May 1;4(4):363-75.
}
\lref\FORM{
	J.A.M.~Vermaseren,
	``New features of FORM,''
	arXiv:math-ph/0010025.
	M.~Tentyukov and J.A.M.~Vermaseren,
	``The multithreaded version of FORM,''
	arXiv:hep-ph/0702279.
}
\lref\gersten{
	M.~Gerstenhaber, ``Developments from Barr's thesis.''
	Journal of Pure and Applied Algebra 143, no. 1-3 (1999): 205-220.
}
\lref\giaquinto{
	A.~Giaquinto, ``Topics in algebraic deformation theory.''
	In Higher structures in geometry and physics, pp. 1-24. Birkh\"auser, Boston, MA, 2011.
}
\lref\garsia{
	A.M. Garsia, ``Combinatorics of the Free Lie Algebra and the Symmetric Group'',
	In Analysis, et Cetera, edited by Paul H. Rabinowitz and Eduard Zehnder,
	Academic Press, (1990) 309-382
}
\lref\garsiareut{
       A.M.~Garsia, C.~Reutenauer, ``A decomposition of Solomon's descent algebra''.
       Advances in Mathematics, 77(2) (1989), 189-262
}
\lref\oneloopI{
	C.R.~Mafra and O.~Schlotterer,
	``Towards the n-point one-loop superstring amplitude. Part I. Pure spinors and superfield kinematics,''
	JHEP \textbf{08}, 090 (2019)
	[arXiv:1812.10969].
}
\lref\oneloopII{
	C.R.~Mafra and O.~Schlotterer,
	``Towards the n-point one-loop superstring amplitude. Part II. Worldsheet functions and their duality to kinematics,''
	JHEP \textbf{08}, 091 (2019)
	[arXiv:1812.10970].
}
\lref\EOMbbs{
	C.R.~Mafra and O.~Schlotterer,
  	``Multiparticle SYM equations of motion and pure spinor BRST blocks,''
	JHEP {\bf 1407}, 153 (2014).
	[arXiv:1404.4986].
}
\lref\partI{
	C.R.~Mafra and O.~Schlotterer,
  	``Cohomology foundations of one-loop amplitudes in pure spinor superspace,''
	[arXiv:1408.3605].
}
\lref\garsiaremmel{
	A.M.Garsia and J.Remmel,
	``Shuffles of permutations and the Kronecker product,''
	Graphs Combin. {\bf 1} (1985), 217-263.
}
\lref\atkinson{
	Atkinson, M. D.
	``Solomon's Descent Algebra Revisited,''
	Bulletin of the London Mathematical Society
	Bull London Math Soc (1992)
}
\lref\bgsym{
	F.A.~Berends and W.T.~Giele,
  	``Multiple Soft Gluon Radiation in Parton Processes,''
	Nucl.\ Phys.\ B {\bf 313}, 595 (1989).
}
\lref\Gauge{
	S.~Lee, C.R.~Mafra and O.~Schlotterer,
  	``Non-linear gauge transformations in $D=10$ SYM theory and the BCJ duality,''
	JHEP {\bf 1603}, 090 (2016).
	[arXiv:1510.08843].
}
\lref\BGBCJ{
	C.R.~Mafra and O.~Schlotterer,
  	``Berends-Giele recursions and the BCJ duality in superspace and components,''
	JHEP {\bf 1603}, 097 (2016).
	[arXiv:1510.08846 [hep-th]].
}

\lref\MSSI{
	C.R.~Mafra, O.~Schlotterer and S.~Stieberger,
  	``Complete N-Point Superstring Disk Amplitude I. Pure Spinor Computation,''
	[arXiv:1106.2645].
}
\lref\MSSII{
	C.R.~Mafra, O.~Schlotterer and S.~Stieberger,
  	``Complete N-Point Superstring Disk Amplitude II. Amplitude and Hypergeometric Function Structure,''
	[arXiv:1106.2646].
}
\lref\singleST{
	S.~Stieberger and T.~R.~Taylor,
	``Closed String Amplitudes as Single-Valued Open String Amplitudes,''
	Nucl. Phys. B \textbf{881}, 269-287 (2014)
	[arXiv:1401.1218 [hep-th]].
}
\lref\datamine{
	J.~Blumlein, D.~J.~Broadhurst and J.~A.~M.~Vermaseren,
  	``The Multiple Zeta Value Data Mine,''
	Comput.\ Phys.\ Commun.\  {\bf 181}, 582 (2010).
	[arXiv:0907.2557 [math-ph]].
}

\listtoc
\writetoc
\filbreak

\newsec Introduction

It is well-known that superstring $n$-point color-ordered disk amplitudes
satisfy monodromy relations which imply that the number of linearly independent amplitudes is
$(n-3)!$, for all $\ap$ corrections \refs{\monodVanhove,\monodStieberger}. These relations involve
coefficients that depend on
Mandelstam variables \mizeraKLT\ and are famously related to the Bern-Carrasco-Johansson (BCJ) color-kinematics
amplitude relations in the field-theory limit. In this paper, we investigate a weaker set of
relations, called KK-like
relations \copenhagen, of higher $\ap$ corrections to disk amplitudes \motivic\ refined by
their motivic MZV
content written in the $f$-alphabet of \brownmot. More precisely, writing the motivic superstring color-ordered disk amplitude
as\foot{For more information on motivic amplitudes and the $f$-alphabet expansion, see
\refs{\motivic,\brownmot}.}
\eqn\AstringMZVintro{
\phi\big(\Astring(1,2, \ldots,n)\big) = \AYM(1,2, \ldots,n) + \sum_{m=0}^\infty\sum_M A_{\z_2^m\z_M}(1,2, \ldots,n)f_2^m f_M
}
where $M$ runs over all words composed of odd positive integers $\{2k+1\mid k\ge 1\}$ and
$f_M = f_{m_1}f_{m_2} \ldots f_{m_p}$ for words of length $p$ (e.g. $f_{3,5,3} = f_3f_5f_3$),
we will focus on relations obeyed by the {\it motivic MZV amplitudes} $A_{\z_2^m \z_M}$ of the form
\eqn\kkdefintro{
\sum_\s c_\s A_{\z_2^m \z_M}(\s) = 0,
}
where the coefficients $c_\s$ are rational numbers independent of Mandelstam invariants.
In addition, the map $\phi$ in \AstringMZVintro\ converts the motivic MZVs to the $f$-alphabet as
described in \refs{\brownmot,\motivic}.
It is well-known that \AstringMZVintro\ is cyclically symmetric and satisfies
$\Astring(1,2, \ldots,n) = (-1)^n \Astring(n,n-1, \ldots,1)$, so
the basis dimensions of motivic MZV amplitudes is at most $\half(n-1)!$.
When the superstring amplitude \AstringMZVintro\ is restricted to its field-theory limit $\AYM$, the
KK-like relations correspond to the famous
Kleiss-Kuijf (KK) relations \KKref, under which there are only $(n-2)!$ independent amplitudes. However,
it is not
yet known the general form of the KK-like relations and the corresponding basis dimensions
for the $\z_2^m\z_M$ components of \AstringMZVintro\ with $m\ge2$. For instance, a brute-force
search indicates that the upper bound $\half(n-1)!$ is saturated for $A_{\z_2^2\z_M}(1,2, \ldots,n)$
when $n=4,5,6$ and $7$, but the $n=8$ case is different. In fact,
\eqn\exsintro{
\#\big(A_{\z_2^2\z_M}(1,2, \ldots,8)\big) = 2519 = \half\,7! - 1.
}
In this paper, this result and its generalization will be obtained as
\eqnn\summdim
$$\eqalignno{
\#\Big(A_{\z_M}(1,2, \ldots,n)\Big) & = {n-1\stirling 1}=(n-2)!&\summdim\cr
\#\Big(A_{\z_2\z_M}(1,2, \ldots,n)\Big) & = {n-1\stirling 3}\cr
\#\Big(A_{\z_2^m\z_M}(1,2, \ldots,n)\Big) & = {n-1\stirling 1} + {n-1\stirling 3}
+ \cdots + {n-1\stirling 2m+1},\quad m\ge2\,.
}$$
We will see that the general KK-like
relations are closely related to the mathematical framework of the Solomon descent algebra \refs{\mackey,\garsiareut,\PBWReutenauer,\schockerSolomon,\garsiaremmel,\thibon,\Reutenauer}.
To see this we will define
the {\it color-dressed permutation}
\eqn\cdressedPintro{
P_{n} = \sum_{\s\in S_{n}, \s(1)=1} \Tr(T^\s)\, \s\,,\qquad T^\s:=T^{\s(1)}T^{\s(2)}\cdots T^{\s(n)}
}
which is inspired by the expression of the color-dressed disk amplitudes, where $T^i:=T^{a_i}$
denotes a Chan-Paton factor.
When the closed formula from \ctrace\ for the color-trace decomposition \Vcolor\ is plugged into \cdressedPintro,
the permutations appearing as coefficients with respect to a basis of color factors define what we call
{\it BRST invariant permutations} $\g_{1|P_1, \ldots,P_k}$
with $1\le k\le n{-}1$. We conjecture the following closed formula
\eqn\closedgamma{
\gamma_{1|P_1, \ldots,P_k}
= 1.\cE(P_1)\shuffle\cE(P_2)\shuffle \ldots \shuffle\cE(P_k)
}
where $\cE(P)$ satisfying $\cE(R\shuffle S)=0$ for $R,S\neq\emptyset$ is the {\it Berends-Giele
idempotent}, defined in section~\BGeulersec\ from mapping the permutations of the Solomon idempotent \solomon\ into their
inverses. Then in section~\correspsec\ we will find evidence that these BRST-invariant permutations encode the
general KK-like relations as
\eqnn\AYMlikeintro
$$\eqalignno{
A_{\z_M}(\g_{1|P_1, \ldots,P_k}) &= 0,\quad k\neq1,  &\AYMlikeintro\cr
A_{\z_2\z_M}(\g_{1|P_1, \ldots,P_k}) &= 0,\quad k\neq3\,\cr
A_{\z_2^m\z_M}(\g_{1|P_1, \ldots,P_k}) &= 0,\quad k\neq1,3,5, \ldots,2m+1,\quad m\ge2.
}$$
Studying the color-dressed string disk amplitude we will obtain, following
the results of \oneloopbb,
a correspondence between the above
permutations and kinematics from the string disk amplitudes. More precisely,
the duality from \oneloopbb\ is generalized to
\eqn\dualintro{
\AYM(1,2, \ldots,n)\;\; \longleftrightarrow\;\; \gamma^{(1)}_{123 \ldots n}\,,\qquad
\AFq(1,2, \ldots,n)\;\; \longleftrightarrow\;\; \gamma^{(3)}_{123 \ldots n}\,,
}
where $\gamma^{(1)}$ and $\gamma^{(3)}$ are orthogonal idempotents of the (inverse) descent algebra constructed
from linear combinations of $\g_{1|P_1, \ldots,P_k}$ with $k=1$ and $k=3$ respectively.
This interpretation is important because we can use a theorem from the work of Garsia and Reutenauer \garsiareut\
to prove results conjectured in \oneloopbb.

In section~\Cgamsec, the BRST-invariant permutations will be used to define higher-mass BRST-invariant
superfields via
\eqn\correspintro{
\g_{1|P_1,P_2, \ldots,P_k} \leftrightarrow \Astring(\g_{1|P_1,P_2, \ldots,P_k})\,,\qquad
A_{\z_2^m \z_M}(\g_{1|P_1, \ldots,P_k}) = k! C^{\z_2^m\z_M}_{1|P_1, \ldots,P_k}\,.
}
As shown in section~\CAsFq, the case for $\z_2$ reduces to the BRST-invariants studied in \refs{\oneloopbb,\EOMbbs}.
In the appendices we review the descent algebra and collect various proofs and explicit expansions
omitted from the main text.

\newsubsubsec\convsec Conventions

Words from the alphabet ${\Bbb N} = \{1,2, \ldots\}$ will be denoted either by capital Latin letters or,
especially when viewed as elements of the permutation group, by lower
case Greek letters.
The symmetric group $S_n$ acts on words of length $n$
via the right-action multiplication \Reutenauer
\eqn\rightaction{
P\circ \si = p_{\si(1)}p_{\si(2)} \ldots p_{\si(n)},
}
where $p_i$ denotes the $i$th letter of $P$.
For example $abcd\circ 3124 = cabd$. The inverse $\s^{-1}$ of a permutation $\s$ of length $n$ is such that
$\s\circ\s^{-1}=\s^{-1}\circ\s = 12 \ldots n$. For example, $(2314)^{-1}=3124$.
For typographical convenience, we write a permutation $\s$
as $W_{\s}$.

\newnewsec\cdressedsec The combinatorics of color-dressed permutations

\newsubsec\cdresseddef Color-dressed permutations

In this section we will investigate the combinatorics of the {\it color-dressed permutations} $P_{n}$
\eqn\cdressedP{
P_{n} = \sum_{\s\in S_{n},\s(1)=1} \Tr(T^\s)\,W_{\s}\,,\qquad T^{\s} := T^{\s(1)}T^{\s(2)}\cdots T^{\s(n)}\,,
}
arising from
decomposing \Vcolor\ the traces of color factors into symmetrized
traces $d^{12 \ldots k}$ and structure constants $f^{abc}$ of the gauge group \ctrace
\eqn\closedformula{
\Tr(T^0T^1\cdots T^{n-1}) = \sum_{S_{n-1}\ni \sigma = \sigma_1\cdots
\sigma_k} i^{n-1-k} \kappa_{\sigma_1} \cdots \kappa_{\sigma_k} d^{0a_1\cdots a_k} F^{\sigma_1}_{a_1}\cdots F^{\sigma_k}_{a_k}.
}
where $\sigma=\s_1\cdot \s_2 \cdot \ldots \cdot \s_k$ denotes the decreasing Lyndon factorization of $\sigma$
to be defined below,
$F^{\s}_a$ for a word $\s$ and a letter $a$ is defined recursively by
$F^{Pj}_a = F^P_b f^{bja}$ with base case $F^{i}_a = \d^i_a$.
The coefficients $\kappa_\s$ are defined in \eulsi. In addition, the
symmetrized trace and the structure constant are given by
\eqn\ddef{
d^{12 \ldots k}:={1\over k!}\sum_{\s \in S_k}\Tr(T^\s),\qquad d^{12}:=\half \d^{12},\quad
[T^a,T^b]=if^{abc}T^c.
}
The {\it decreasing Lyndon factorization} (dLf) of $\s$ is defined as \refs{\lothaire,\garsiareut}
$\s=\s_1.\s_2 \ldots \s_k$
representing
the unique
deconcatenation of $\sigma$ into Lyndon subwords
$\sigma_1,\ldots ,\sigma_k$ such that $\sigma_1>\cdots>\sigma_k$ in the lexicographical order of the alphabet ${\Bbb N}=\{1,2,
\ldots,\}$.
Representing the concatenation by a dot to distinguish the subwords $\s_i$ in the dLF factorization of $\s$,
we have, for example
$1432 = 1432$, $2134 = 2.134$, $54132 = 5.4.132$, $42671835 = 4.267.1835$.

\proclaim Definition (BRST-invariant permutations).
The {\it BRST-invariant permutations} are the coefficients with respect to
the basis of color factors
$d^{1a_1 \ldots a_k}F^{\s_1}_{a_1} \ldots F^{\s_k}_{a_k}$, where
$\s_1 \cdots\s_k$ is the decreasing Lyndon factorization of $\s$,
in the
color-dressed permutation \cdressedP, i.e.
\eqn\BRSTdef{
P_{n} = \sum_{\sigma\in S_{n-1}}i^{n-1-k}
d^{1a_1 \ldots a_k}F^{\s_1}_{a_1} \cdots F^{\s_k}_{a_k}\,\gamma_{1|\s_1, \ldots,\s_k}\,,
}\par
\noindent The reason for this terminology will become clear in \gamseries\ when $\g_{1|\s_1,\s_2,\s_3}$ will
be associated to BRST invariants superfields in the pure spinor formalism.
\noindent For example,
plugging
$\Tr(T^1 T^2 T^3) = d^{1a_1a_2}F^2_{a_1}F^{3}_{a_2} + {i\over2}d^{1a}F^{23}_a$
into \cdressedP\ yields
\eqn\Mthree{
P_3 = d^{1ab}F^2_{a}F^3_{b}\g_{1|2,3} + i d^{1a}F^{23}_{a}\g_{1|23}\,,
}
with
\eqn\multthreega{
\gamma_{1|2,3} = W_{123} + W_{132}\,,\quad
\gamma_{1|23} ={1\over2}W_{123} - {1\over2}W_{132}\,.
}
Repeating the same exercise for $n=4$ using \closedformula\
we obtain
\eqnn\linearcombs
$$\eqalignno{
P_4 &= d^{1abc}F^2_aF^3_bF^4_c\,  \gamma_{1|2,3,4} &\linearcombs\cr&
       + i d^{1ab} F^{23}_aF^4_b\, \gamma_{1|23,4}
       + i d^{1ab} F^{24}_aF^3_b\, \gamma_{1|24,3}
       + i d^{1ab} F^{2}_aF^{34}_b\, \gamma_{1|2,34}\cr&
       +i^2 d^{1a}F^{234}_a\, \gamma_{1|234}
       +i^2 d^{1a}F^{243}_a\, \gamma_{1|243}
}$$
where the BRST-invariant permutations are given by
\eqnn\multfourgammas
$$\eqalignno{
\gamma_{1|2,3,4} &=
	    W_{1234}
          + W_{1243}
          + W_{1324}
          + W_{1342}
          + W_{1423}
          + W_{1432}\,, &\multfourgammas\cr
\gamma_{1|23,4} &=
	    {1\over2} W_{1234}
          + {1\over2} W_{1243}
          - {1\over2} W_{1324}
          - {1\over2} W_{1342}
          + {1\over2} W_{1423}
          - {1\over2} W_{1432}\,,
\cr
\gamma_{1|2,34} &=
	   {1\over2} W_{1234}
          - {1\over2} W_{1243}
          + {1\over2} W_{1324}
          + {1\over2} W_{1342}
          - {1\over2} W_{1423}
          - {1\over2} W_{1432}\,,\cr
\gamma_{1|24,3} &=
	   {1\over2} W_{1234}
          + {1\over2} W_{1243}
          + {1\over2} W_{1324}
          - {1\over2} W_{1342}
          - {1\over2} W_{1423}
          - {1\over2} W_{1432}\,,\cr
\gamma_{1|234} &=
	    {1\over3} W_{1234}
          - {1\over6} W_{1243}
          - {1\over6} W_{1324}
          - {1\over6} W_{1342}
          - {1\over6} W_{1423}
          + {1\over3} W_{1432}\,,\cr
\gamma_{1|243} &=
	  - {1\over6} W_{1234}
          + {1\over3} W_{1243}
          - {1\over6} W_{1324}
          + {1\over3} W_{1342}
          - {1\over6} W_{1423}
          - {1\over6} W_{1432}\,.
}$$
For $n=5$ we obtain
\eqnn\Mfive
$$\eqalignno{
P_5 &= d^{1abcd}F^2_aF^3_bF^4_cF^5_d\, \g_{1|2,3,4,5} &\Mfive\cr&
+ i d^{1abc}F^{23}_a F^4_b F^5_c\, \g_{1|23,4,5}
+ i d^{1abc}F^{24}_a F^3_b F^5_c\, \g_{1|24,3,5}
+ i d^{1abc}F^{25}_a F^3_b F^4_c\, \g_{1|25,3,4}\cr&
+ i d^{1abc}F^{2}_a F^{34}_b F^5_c\, \g_{1|2,34,5}
+ i d^{1abc}F^{2}_a F^{35}_b F^4_c\, \g_{1|2,35,4}
+ i d^{1abc}F^{2}_a F^3_b F^{45}_c\, \g_{1|2,3,45}\cr&
+ i^2 d^{1ab}F^{234}_a F^5_b\, \g_{1|234,5}
+ i^2 d^{1ab}F^{243}_a F^5_b\, \g_{1|243,5}
+ i^2 d^{1ab}F^{235}_a F^4_b\, \g_{1|235,4}\cr&
+ i^2 d^{1ab}F^{245}_a F^3_b\, \g_{1|245,3}
+ i^2 d^{1ab}F^{253}_a F^4_b\, \g_{1|253,4}
+ i^2 d^{1ab}F^{254}_a F^3_b\, \g_{1|254,3}\cr&
+ i^2 d^{1ab}F^{345}_a F^2_b\, \g_{1|345,2}
+ i^2 d^{1ab}F^{354}_a F^2_b\, \g_{1|354,2}\cr&
+ i^2 d^{1ab}F^{23}_a F^{45}_b\, \g_{1|23,45}
+ i^2 d^{1ab}F^{24}_a F^{35}_b\, \g_{1|24,35}
+ i^2 d^{1ab}F^{25}_a F^{34}_b\, \g_{1|25,34}\cr&
+ i^3 d^{1a}F^{2345}_a\, \g_{1|2345}
+ i^3 d^{1a}F^{2354}_a\, \g_{1|2354}
+ i^3 d^{1a}F^{2435}_a\, \g_{1|2435}\cr&
+ i^3 d^{1a}F^{2453}_a\, \g_{1|2453}
+ i^3 d^{1a}F^{2534}_a\, \g_{1|2534}
+ i^3 d^{1a}F^{2543}_a\, \g_{1|2543}
}$$
where the various $\g_{1|A_1,A_2, \ldots,A_k}$
are listed
in the appendix~\higherap.

\newsubsubsec\motivsec Relating the BRST-invariant permutations with the descent algebra

The permutations in each $\g_{1|A_1,A_2, \ldots,A_k}$ turn out to be
related
to the descent algebra reviewed in the appendix~\SolomonDSsec. To see this
consider $\g_{1|23,4}$ from \multfourgammas,
relabel $i\to i-1$ and strip off the leading ``0'' (denoted by $\times$) to obtain
\eqn\notinSigma{
\gamma_{\times|12,3} =
	    {1\over2} \underbrace{W_{123}}_{\in D_\emptyset}
          + {1\over2} \underbrace{W_{132}}_{\in D_{\{2\}}}
          - {1\over2} \underbrace{W_{213}}_{\in D_{\{1\}}}
          - {1\over2} \underbrace{W_{231}}_{\in D_{\{2\}}}
          + {1\over2} \underbrace{W_{312}}_{\in D_{\{1\}}}
          - {1\over2} \underbrace{W_{321}}_{\in D_{\{1,2\}}}\,,
}
The above permutations
are {\it not} in the descent algebra ${\cal D}_3$ since permutations in the same
descent class have different coefficients (see also proposition 2.1 from \BGschocker).
However, the inverse permutations in $\t(\g_{\times|12,3})$ do
belong to the same descent classes:
\eqnn\inSigma
$$\eqalignno{
\t\big(\gamma_{\times|12,3}\big) &=
	    {1\over2} \underbrace{W_{123}}_{\in D_\emptyset}
          + {1\over2} \underbrace{W_{132}}_{\in D_{\{2\}}}
          - {1\over2} \underbrace{W_{213}}_{\in D_{\{1\}}}
          - {1\over2} \underbrace{W_{312}}_{\in D_{\{1\}}}
          + {1\over2} \underbrace{W_{231}}_{\in D_{\{2\}}}
          - {1\over2} \underbrace{W_{321}}_{\in D_{\{1,2\}}} &\inSigma\cr
	  & = \half D_\emptyset - \half D_{\{1\}} + \half D_{\{2\}} - \half D_{\{1,2\}}\,.
}$$
as can be verified using \Sthree.
This suggests that the BRST-invariant permutations belong
to the {\it inverse} descent algebra ${\cal D}_n':=\t({\cal D}_n)$. To find an algorithm that
generates these permutations, we will
consider the inverse of the
Eulerian idempotent \eulsi.

\newsubsec\BGeulersec The Berends-Giele idempotent

Define the {\it Berends-Giele idempotent}\foot{The inverse of an idempotent is also idempotent.}
$\cE_n$ as the inverse of the Eulerian idempotent \eulsi:
\eqn\BGeuler{
\cE_n = \sum_{\s\in S_n}\kappa_{\s^{-1}}\s\,,\qquad
\cE(P) = \cE_P := P\circ\cE_n\,,\quad|P|=n\,.
}
The reason for this terminology is the
correspondence with the standard Berends-Giele current of
Yang-Mills theory \BerendsME, see section~\AFqsec.
A few examples of \BGeuler\ are
\eqnn\cEexs
$$\displaylines{
\cE(1) = W_1,\qquad \cE(12) = \half(W_{12}-W_{21})\,, \hfil\cEexs\hfilneg\cr
\cE(123) =
       {1\over3} \Word(1,2,3)
       - {1\over6} \Word(1,3,2)
       - {1\over6} \Word(2,1,3)
       - {1\over6} \Word(2,3,1)
       - {1\over6} \Word(3,1,2)
       + {1\over3} \Word(3,2,1)\,,\cr
}$$
while the expansion of $\cE(1234)$ can be found in the appendix~\BGexapp.

\proclaim Proposition (Shuffle Symmetry). The Berends-Giele idempotent \BGeuler\ satisfies
\eqn\shufband{
\cE(R\shuffle S) = 0\,,\qquad R,S\neq\emptyset\,.
}\par
\noindent{\it Proof.}\foot{We know from \Ree\
that any Lie polynomial can be expanded as $\sum_\s M_\s \s$ with $M_{R\shuffle S}=0$ for nonempty
$R,S$, so it follows
that if $\Gamma$ is a Lie polynomial then the word function ${\cal F}(P):=P\circ\t(\Gamma)$ satisfies
the shuffle symmetry ${\cal F}(R\shuffle S)=0$ for $R,S\neq\emptyset$.
} Since the sum in \BGeuler\ is over all permutations we rename $P\circ\s = \tau$ and sum over
$\tau$. Notice that $\s^{-1} = \tau^{-1}\circ P$, so
$\cE(P) = \sum_\s \kappa_{\s^{-1}} P\circ\s = \sum_\tau \kappa_{(\tau^{-1}\circ P)}\tau$
and therefore
\eqn\funshu{
\cE(R\shuffle S) = \sum_\tau \kappa_{(\tau^{-1}\circ(R\shuffle S))}\tau =
\sum_\tau \kappa_{(\tau^{-1}(R)\shuffle \tau^{-1}(S))}\tau = 0
}
where the last equality follows from  \kapshu\ and the crucial observation in (1.5) of \garsiareut\ that
$\s^{-1}\circ(R\shuffle S)=\s^{-1}(R)\shuffle \s^{-1}(S)$,
where $\s^{-1}(R)$ denotes
the word obtained by replacing each letter in $R$ by its image under $\s^{-1}$. \QED

\newsubsec\newIpsec Inverse idempotent basis and BRST-invariant permutations

Following the realization in section~\motivsec\ that the
BRST-invariant permutations are related to the inverse of the descent algebra,
we define the inverse of the idempotent basis $I_p$ as
\eqn\cIpdef{
\cI_{p_1p_2 \ldots p_k}(P_1,P_2, \ldots,P_k) :=
\theta(I_{p_1p_2 \ldots p_k})(P)\,,\quad \len{P_i}= p_i\,,\quad P_1P_2 \ldots P_k = P
}
where the map $\t$ is defined in \thetamap. For example
$\cI_{21}(12,3) = \half\big(
       \Word(1,2,3)
       + \Word(1,3,2)
       - \Word(2,1,3)
       - \Word(2,3,1)
       + \Word(3,1,2)
       - \Word(3,2,1)\big)$.
See \cIptwotwo\ for the explicit permutations in $\cI_{22}(12,34)$.

An alternative representation is proven in the appendix~\cIpap
\eqn\proptIp{
\cI_{p_1p_2 \ldots p_k}(P_1,P_2, \ldots,P_k)
= \cE(P_1)\shuffle\cE(P_2)\shuffle \ldots\shuffle \cE(P_k)\,,
\quad\len{P_i}=p_i\,,
}
from which it follows that ($\tilde P$ denotes
the reversal\foot{If $Q=q_1 \ldots q_n$ then its reversal is the word $\tilde Q:=q_n \ldots q_1$.}
of $P$)
\eqnn\shuffleIp
\eqnn\revcIp
$$\eqalignno{
\cI_{\ldots p_i \ldots}( \ldots ,R\shuffle S,\ldots) &= 0, \quad R,S\neq\emptyset,\quad \len{R}+\len{S}=p_i\,,&\shuffleIp\cr
\cI_{\ldots p_i\ldots p_j \ldots}(\ldots, P_i,\ldots,P_{j},\ldots)&=
\cI_{\ldots p_j \ldots p_i \ldots }( \ldots, P_j,\ldots,P_{i}, \ldots)\,,\cr
{\tilde\cI}_{p_1 \ldots p_k}(P_1, \ldots,P_k) &= (-1)^{\#{\rm even}(p)}\cI_{p_1 \ldots p_k}(P_1,
\ldots,P_k)\,, &\revcIp
}$$
where $\#{\rm even}(p)$ denotes the number of even parts in the composition $p$.
To see this note that if a function satisfies the shuffle symmetry or, in other words, belongs
to the dual space of Lie polynomials \flas, then $\tilde F(P)=(-1)^{\len{P}-1} F(P)$ and \revcIp\
follows from \proptIp.

Finally, based on multiple examples we conjecture that
\eqn\gammacIp{
\gamma_{1|P_1, \ldots,P_k}
= 1.\cI_{p_1 \ldots p_k}(P_1, \ldots,P_k)
= 1.\cE(P_1)\shuffle\cE(P_2)\shuffle \ldots \shuffle\cE(P_k)\,,\quad p_i=\len{P_i}\,.
}
The shuffle symmetry of $\cE(P_i)$ can be used to fix
the first letter of $P_i$ and the commutativity of the shuffle product implies total symmetry
in word exchanges, so
the basis dimension of $\gamma_{1|P_1, \ldots,P_k}$ with $n$ letters
is given by the Stirling
cycle numbers \tnn
\eqn\dimgamma{
\#\big(\gamma_{1|P_1,P_2, \ldots, P_k}\big) = {n-1\stirling k}\,,\qquad \sum_{i=1}^k\len{P_i} =
n-1\,,
}
which explains the total number of terms in $P_n$ as
$\sum_{k=1}^{n-1}{n-1\stirling k}=(n-1)!$.

\newsubsec\Bdefssec BRST-invariant permutations and orthogonal idempotents

Since the BRST-invariant permutations have been related to the idempotent
basis of the (inverse) descent algebra in \gammacIp\ we may construct orthogonal
idempotents as in section~\orthsec\ by taking the inverse of the Reutenauer idempotents \Ereutdef\
$\g^{(i)}_{12 \ldots n} := 1.\theta\big(E^{(i)}\big)$,
where the labels in $\theta\big(E^{(i)}\big)$ must be shifted as $i\to i+1$
prior to the left concatenation with the letter $1$.
Equivalently, from \Elambdadef, \Ereutdef, and \gammacIp\ we obtain
\eqn\gamieq{
\g^{(k)}_{12 \ldots n} =\!\!\!\!\!\!\! \sum_{23 \ldots n=P_1 \ldots P_k}{1\over k!}\g_{1|P_1, \ldots, P_k}\,.
}
From the discussion of section~\orthsec\ it follows that \gamieq\
are orthogonal idempotents in the inverse descent algebra ${\cal D}'_n$
satisfying ($\d^{ij}$ is the Kronecker delta)
\eqn\fiveorthos{
\sum_{k=1}^{n-1} \g^{(k)}_{12 \ldots n} =W_{12 \ldots n}\,,\qquad
\g^{(i)}_{12 \ldots n}\circ\g^{(j)}_{12 \ldots n} = \d^{ij}\g^{(i)}_{12 \ldots n}
}
For example, from the BRST-invariant permutations in \multthreega\ we get
\eqn\threedefs{
\g^{(1)}_{123}\equiv\g_{1|23}=\half W_{123}-\half W_{132},
\quad \g^{(2)}_{123}\equiv{1\over2}\g_{1|2,3}=\half W_{123} + \half W_{132}\,,
}
which clearly satisfies \fiveorthos.
Similarly, at multiplicity four the definition \gamieq\
\eqn\Amfour{
\g^{(1)}_{1234} = \g_{1|234}\,,\qquad
\g^{(2)}_{1234} = {1\over 2}\big(\g_{1|23,4} + \g_{1|2,34}\big)\,,\qquad
\g^{(3)}_{1234} = {1\over 3!}\g_{1|2,3,4}\,,
}
yields
\eqnn\fourptga
$$\eqalignno{
\g^{(1)}_{1234} &=
            {1\over3} W_{1234}
          - {1\over6} W_{1243}
          - {1\over6} W_{1324}
          - {1\over6} W_{1342}
          - {1\over6} W_{1423}
          + {1\over3} W_{1432}\,,&\fourptga\cr
\g^{(2)}_{1234} &=
            {1\over2} W_{1234}
          - {1\over2} W_{1432}\,,\cr
\g^{(3)}_{1234} &=
            {1\over6} W_{1234}
          + {1\over6} W_{1243}
          + {1\over6} W_{1324}
          + {1\over6} W_{1342}
          + {1\over6} W_{1423}
	  + {1\over6} W_{1432}\,.
}$$
It is straightforward but tedious \FORM\ to check that the above satisfy \fiveorthos.
At multiplicity five, the orthogonal idempotents are given by
\eqn\Amfive{
\eqalign{
\g^{(1)}_{12345} &= \g_{1|2345}\,,\cr
\g^{(2)}_{12345} &= {1\over2}\big(\g_{1|234,5} + \g_{1|23,45} + \g_{1|2,345}\big)\,,
}\quad
\eqalign{
\g^{(3)}_{12345} &= {1\over 3!}\big(\g_{1|23,4,5} + \g_{1|2,34,5} +\g_{1|2,3,45}\big)\,,\cr
\g^{(4)}_{12345} &= {1\over 4!}\g_{1|2,3,4,5}\,,\cr
}
}
whose expansions can be found in the appendix~\higherap\ and can be checked to obey \fiveorthos.

\newnewsec\descKKsec The descent algebra and string disk amplitudes

\newsubsec\correspsec KK-like relations of $\ap$ corrections to disk amplitudes

The color-dressed string motivic disk amplitude
\eqn\cdressedS{
M_n = \!\!\!\!\sum_{\s \in S_{n-1}}\!\!\!
\Tr
\bigl( T^{1} T^{\si(2)} \cdots T^{\si(n)} \bigr)
\, \phi\big(A^{\rm string}(1,\si(2),\ldots,\si(n))\big)\,,
}
is a sum over disk orderings of the ($\phi$ map of) open string color-ordered amplitude weighted by traces of
Chan-Paton factors. The explicit form of the disk amplitudes is a linear combination
of field-theory amplitudes $\AYM$ of ten-dimensional super-Yang-Mills \SYM\
given by \refs{\MSSI,\BGap}
\eqn\AstringP{
A^{\rm string}(P) = \sum_{Q,R \in S_{n-3}} Z(P|1,R,n,n{-}1) S[R|Q]_1 A^{\rm SYM}(1,Q,n{-}1,n)
}
where $S(P|Q)_1$ is the field-theory KLT kernel \refs{\oldMomKer,\BohrMomKer,\momKern}
conveniently computed recursively by $S[A,j|B,j,C]_i =
(k_{iB}\cdot k_{j}) S[A|B,C]_i$, with
$S[\varnothing| \varnothing]_i := 1$ \refs{\NLSM,\PScomb,\flas}. In addition,
$Z(P|Q)$ are the non-abelian $Z$-theory amplitudes of \refs{\Polylogs,\BGap}.
The color-ordered motivic amplitudes \motivic\ are decomposed as
\eqn\AstringMZV{
\phi\big(\Astring(1,2, \ldots,n)\big) = \AYM(1,2, \ldots,n) + \sum_{m=0}^\infty\sum_M
A_{\z_2^m\z_M}(1,2, \ldots,n)f_2^m f_M
}
where $M$ runs over all words composed of odd positive integers ${\ge3}$ and $f_M=f_{m_1}f_{m_2} \ldots
f_{m_p}$ in the $f$-alphabet of \brownmot\ for words of length $p$.
From now on we will use the shorthand {\it motivic MZV
amplitudes} for the
$A_{\z_2^m \z_M}$ components of \AstringMZV, and we will see below that components with the same
$\ap$ but different motivic MZV content satisfy different relations.

It is well-known that the color-ordered string disk amplitudes are cyclically symmetric and parity
invariant \schwarzrev
\eqnn\symmetries
$$\eqalignno{
\Astring(1,2, \ldots,n) &= \Astring(2,3, \ldots,n,1) &\symmetries\cr
\Astring(1,2, \ldots,n) &=(-1)^n\Astring(n,n-1, \ldots,2,1)\,.
}$$
These relations imply an upper bound on the number of linearly-independent motivic MZV amplitudes $A_{\z_2^m
\z_M}(1,2, \ldots,n)$
of $\half(n-1)!$. We want to know whether the different motivic MZV amplitudes in \AstringMZV\ satisfy
additional {\it KK-like} relations and their corresponding basis dimensions, where:
\proclaim Definition (KK-like relation). An identity of the form
\eqn\kklikedef{
\sum_\s c_\s A_{\z_2^m \z_M}(\s) = 0,
}
where the coefficients $c_\s$ are rational numbers is said to be KK-like \copenhagen.\par

\noindent For an example KK-like relation, one can verify that the $\z_2$ amplitudes satisfy \oneloopbb,
\eqn\kkzetatwo{
A_{\z_2}(1,2,3,4,5) + {\rm perm}(2,4,5) = 0\,.
}
A central observation of this paper is that the BRST-invariant permutations $\g_{1|P_1,
\ldots,P_k}$ seem to encode all the KK-like relations among the motivic MZV amplitudes.
While the cyclicity relation in \symmetries\ was used in the definition \cdressedS\ to fix the
label in $\g_{1| \ldots}$, the parity and cyclicity relations \symmetries\ are
encoded by compositions of $n{-}1$ with even number of parts:
\eqn\paritygam{
\Astring(\g_{1|P_1, \ldots,P_k}) = 0,\quad \hbox{$k$ even}.
}
See the appendix~\diskparityap\ for a proof. The identity $\sum_{k\; \rm even}{n-1\stirling k} =
\half(n-1)!$ agrees with the counting of linearly independent amplitudes following from \symmetries.
Note that \paritygam\ implies the parity relation in \symmetries\ as
\eqn\sumpar{
\sum_{k\;\rm even}{1\over k!}\Astring(\g_{1|P_1, \ldots,P_k}) = \Astring(1,2, \ldots,n) -
(-1)^n\Astring(1,n,n{-}1, \ldots,2)\,.
}
Based on explicit computations, we
find that motivic MZV amplitudes
with different $\z_2^m$ content\foot{
The string monodromy relations \refs{\monodVanhove,\monodStieberger} give rise to deformations of the field-theory BCJ relations by powers of
$\ap^{2m}\z_2^m$ \refs{\NLSM,\broedeldixon}.
Since the BCJ-satisfying $\z_M$ components
lead to the minimum $(n-3)!$ dimension, they are not
expected to modify the dimensions within a given $\z_2^m\z_M$ class.} satisfy the following KK-like relations:
\eqnn\AYMlike
$$\eqalignno{
A_{\z_M}(\g_{1|P_1, \ldots,P_k}) &= 0,\quad k\neq1,  &\AYMlike\cr
A_{\z_2\z_M}(\g_{1|P_1, \ldots,P_k}) &= 0,\quad k\neq3\,\cr
A_{\z_2^m\z_M}(\g_{1|P_1, \ldots,P_k}) &= 0,\quad k\neq1,3,5, \ldots,2m+1,\quad m\ge2,
}$$
which constitute the {\it descent algebra decomposition of KK-like relations}.
To count the basis dimensions implied by \AYMlike\ we recall that
$\#\big(\g_{1|P_1, \ldots,P_k}\big) = {n-1\stirling k}$ and $\sum_{k=1}^{n-1}{n-1\stirling k} =
(n-1)!$. Thus, subtracting the dimensions of the BRST-invariant permutations from the number of cyclically
symmetric $n$-point amplitudes\foot{Note that $\half(n-1)!$
is equal to the sum over ${n-1\stirling k}$ with even $k$. Since $k$ even is included in
\AYMlike, the Stirling cycle numbers are subtracted from $(n-1)!$.}
leads to
\eqnn\dofAYMag
\eqnn\dofAFqag
\eqnn\dimzetaeven
$$\eqalignno{
\#\big(A_{\z_M}(\g_{1|P_1, \ldots,P_k})\big) &={n-1\stirling 1}= (n-2)!\,, &\dofAYMag\cr
\#\big(A_{\z_2\z_M}(\g_{1|P_1, \ldots,P_k})\big) &= {n-1\stirling 3}\,,&\dofAFqag\cr
\#\big(A_{\z_2^m\z_M}(\g_{1|P_1, \ldots,P_k})\big) &= {n-1\stirling 1} + {n-1\stirling 3}
+ \cdots +{n-1\stirling 2m+1}\,. &\dimzetaeven
}$$
The dimension \dofAYMag\ corresponds to the number of independent amplitudes under the Kleiss-Kuijf
relations $\AYM(P1Q,n) - (-1)^\len{P}\AYM(1,\tilde P\shuffle Q,n) = 0$, which is also valid for
the BCJ-satisfying $\z_M$ corrections. The dimension \dofAFqag\ was obtained in
\oneloopbb\ following a similar discussion of the all-plus one-loop amplitudes of \copenhagen\
(see also \schabinger).

The basis dimension formula \dimzetaeven\ and the corresponding amplitude relations in \AYMlike\
are new. Interestingly, they imply that the basis dimension of the
$\z_2^m\z_M$ components with $m\ge2$ is given by $\half(n-1)!$ only when
$n\le 2m+3$. More explicitly,
\eqn\lesszeta{
\#\big(A_{\z_2^m\z_M}(1,2, \ldots,n)\big) = \half (n-1)!,\quad n=4,5, \ldots,2m+3\,.
}
For a deviation from the $\half(n-1)!$ dimension, we have, for example with $m=2$
\eqn\deviation{
\#\big(A_{\z_2^2\z_M}(1,2, \ldots,8)\big) = {7\stirling1}+{7\stirling3}+{7\stirling5}=2519 = \half7!-1\,,
}
which was confirmed by a long brute-force search using {\tt FORM} \FORM.
The additional relation on top of \paritygam\ is seen to be
$A_{\z_2^2}(\g_{1|2,3,4,5,6,7,8}) = A_{\z_2^2}(1,2\shuffle 3\shuffle \ldots\shuffle8)= 0$, in
agreement with the analysis of \refs{\NLSM,\semiZ}.
At nine points and $m=2$, the formula \dimzetaeven\ predicts a mismatch due to ${8\stirling 7}=28$ additional
relations etc.

\topinsert
\def\dvrule{\vrule\hskip 0.02cm \vrule}
\def\hquad{\hskip0.63em\relax}
\centerline{
\vbox{
\halign{\strut\vrule\hfil\hquad $#$\hquad\hfil &\dvrule\hfil\hquad$#$\hquad\hfil &\dvrule\hfil\hquad$#$\hquad\hfil
\vrule \tabskip=0pt\cr
\noalign{\hrule}
k & \phi\big(A^{\rm string}(\g_{1|P_1, \ldots,P_k})\big) = 0  & \phi\big(A^{\rm string}(\g_{1|P_1,\ldots,P_k})\big)\neq 0\cr
\noalign{\hrule}
7 &  \z_7,\z_5,\z_3,\z_3^2,\z_2,\z_2\z_5,\z_2\z_3,\z_2^2,\z_2^2\z_3 & \z_2^3 \cr
\noalign{\hrule}
6 &  \forall \z_M & \times \cr
\noalign{\hrule}
5 &  \z_7,\z_5,\z_3,\z_3^2,\z_2,\z_2\z_5,\z_2\z_3 & \z_2^2,\z_2^2\z_3,\z_2^3 \cr
\noalign{\hrule}
4 &  \forall \z_M & \times \cr
\noalign{\hrule}
3 &  \z_7,\z_5,\z_3,\z_3^2 & \z_2,\z_2\z_5,\z_2\z_3,\z_2^2,\z_2^2\z_3,\z_2^3 \cr
\noalign{\hrule}
2 &  \forall \z_M & \times \cr
\noalign{\hrule}
1 &  \z_2,\z_2\z_3,\z_2\z_5 & \z_7,\z_5,\z_3,\z_3^2,\z_2^2,\z_2^2\z_3,\z_2^3 \cr
\noalign{\hrule}
}}
}
\smallskip
{\leftskip=0pt\rightskip=20pt\baselineskip6pt\noindent\ninepoint
{\bf Table 1.} Overview of the descent algebra symmetries of higher $\ap$ corrections to
string disk amplitudes of up to $n=8$ points displayed by their
motivic MZV content of weight $w \leq 7$. The entries
depend only on the number of parts $k$ of the composition of $n{-}1$.
However, a partition with $k$ parts cannot be probed by disk amplitudes with fewer than $k{+}1$ points.
\par}
\endinsert

The evidence for the descent algebra decomposition of KK-like relations in \AYMlike\
was collected from explicit calculations of $A_{\z_2^n\z_M}(\g_{1|P_1, \ldots,P_k})$
for various compositions $p=(p_1, \ldots,p_k)\models n-1$ with $\len{P_i}=p_i$.
The results are summarized in Table 1\foot{This data was collected using the $\ap$ corrections to disk
amplitudes
obtained in \refs{\MSSI,\MSSII,\drinfeld,\BGap} (see also
\refs{\medina,\oprisa,\multigluon,\ragoucy} and references therein for earlier work and \datamine\
for a discussion on MZVs).}.
The number of
checks of $A_{\z_2^m\z_M}(\g_{1|P_1, \ldots,P_k})$ at $n$ points would appear to grow
rapidly, but luckily
a vanishing outcome was observed to depend only
on the number of parts $k$ of the composition, independently of $n$ (with data up to $n=8$).
Thus it suffices to test the single case $k=n{-}1$ for each additional external leg.
Since $\g_{1|2,3, \ldots,n} = 1.(2\shuffle 3\shuffle \ldots \shuffle n)$ by \gammacIp,
we get a
sum over all cyclic orderings of the $n$-point string disk amplitude \AstringP, leading to the
$\ap$-corrected abelian $Z$-theory amplitudes $\Astring(\g_{1|2,3, \ldots,n}) \sim A^{\rm
NLSM}(1,2, \ldots,n)$ of \refs{\NLSM,\semiZ}.
As a consistency check,
the proof \propeven\ implying that $A_{\z_2^m\z_M}(\g_{1|2, \ldots,k})$
vanishes when $k$ is even
(so $n$ is odd) agrees with the vanishing of NLSM odd-point amplitudes.

\newsubsec\AFqsec The field-theory and $\ap^2$ corrections

The SYM amplitudes are computed in pure spinor superspace from the expression $ M_1E_P$,
where $E_P$ is a superfield satisfying the same shuffle symmetry $E_{R\shuffle
S}=0$ for  $R,S\neq\emptyset$ of the standard Berends-Giele current $J^m_P$ \refs{\bgsym,\Gauge}.
The $A_{\z_2}$ amplitudes \greenteight\ can be computed in pure spinor superspace \oneloopbb\ using
BRST-closed combinations of superfields $C_{1|X,Y,Z}$ symmetric
under exchanges of any pairs $X\leftrightarrow Y,Z$ and satisfying
$C_{1|R\shuffle S,Y,Z} = 0$ for $R,S\neq\emptyset$.
For convenience, define the BRST-closed combination $C_{1|P} := M_1 E_P$
so that\foot{The angular brackets $\langle \ldots\rangle$ denotes the pure spinor zero-mode
integration of \psf, but it plays no role in the subsequent discussions.} 
\eqnn\AymC
\eqnn\Afq
$$\eqalignno{
\AYM(1,2, \ldots,n)&=\langle C_{1|2 \ldots n}\rangle\,, &\AymC\cr
A_{\z_2}(1,2, \ldots,n) &= \sum_{2 \ldots n}\langle C_{1|X,Y,Z}\rangle\,,&\Afq
}$$
where $\sum_{2 \ldots n}$ is a shorthand for the sum over the deconcatenations
of $2 \ldots n=XYZ$.
In terms of these BRST invariants, the color-dressed amplitudes at four and five points
can be written as \oneloopbb
\eqnn\fourdr
$$\eqalignno{
M_4 &=
6\z_2 \ap^2  d^{1234} \, C_{1|2,3,4} &\fourdr\cr
&+ i^2 d^{1a}F^{234}_a C_{1|234}
+ i^2 d^{1a}F^{243}_a C_{1|243}
  +  {\cal O}(\ap^3)\,,\cr
M_5 &= 6\z_2\ap^2\,id^{1abc}\Bigl(F^{23}_a F^4_b F^5_c\, C_{1|23,4,5}
+ F^{24}_a F^3_b F^5_c\, C_{1|24,3,5}
+ F^{25}_a F^3_b F^4_c\, C_{1|25,3,4}\cr&
\qquad{}+ F^{2}_a F^{34}_b F^5_c\, C_{1|2,34,5}
+ F^{2}_a F^{35}_b F^4_c\, C_{1|2,35,4}
+ F^{2}_a F^3_b F^{45}_c\, C_{1|2,3,45}\Bigr)\cr&
+ i^3 d^{1a}F^{2345}_a\, C_{1|2345}
+ i^3 d^{1a}F^{2354}_a\, C_{1|2354}
+ i^3 d^{1a}F^{2435}_a\, C_{1|2435}\cr&
+ i^3 d^{1a}F^{2453}_a\, C_{1|2453}
+ i^3 d^{1a}F^{2534}_a\, C_{1|2534}
+ i^3 d^{1a}F^{2543}_a\, C_{1|2543}
  +  {\cal O}(\ap^3)\,,
}$$
with similar expansions at higher points.
Comparing \fourdr\ with the color-dressed permutations $P_4$ and $P_5$ in \linearcombs\ and \Mfive\
implies the correspondences\foot{The first correspondence in \Ctogamma\ implies
$M_1 E_P \leftrightarrow  1\cdot \cE_P$
and suggests the duality
$E_P\leftrightarrow \cE_P$ ($E_P$ here is the superfield of \nptMethod, not the Eulerian idempotent). Since
the superfield $E_P$ is related to the Berends-Giele current $M_P$ \nptMethod, this
motivates the terminology of $\cE_P$ in \BGeuler.} at the given $\ap$ order,
\eqn\Ctogamma{
C_{1|X}\;\; \longleftrightarrow\;\; \gamma_{1|X} \,,\qquad
C_{1|X,Y,Z}\;\; \longleftrightarrow\;\; {1\over6}\gamma_{1|X,Y,Z}\,.
}
So the $\AYM$ and $\AFq$ amplitudes correspond to idempotents of the
descent algebra
\eqn\AFqtogamma{
\AYM(1,2, \ldots,n)\;\; \longleftrightarrow\;\; \gamma^{(1)}_{123 \ldots n}\,,\qquad
\AFq(1,2, \ldots,n)\;\; \longleftrightarrow\;\; \gamma^{(3)}_{123 \ldots n}\,.
}
where the deconcatenations \Afq\ and \gamieq\ have been used in the last line.
The correspondences in \AFqtogamma\ can be used to justify the first two
relations in \symmetries\ from the theorem 4.2 of \garsiareut, namely $E_\mu \circ I_p = 0$ if
$\l(p)\neq\mu$,
where $E_\mu$ for a partition $\mu$ is reviewed in \Elambdadef, $p$ is a composition and $\l(p)$
is its shape as defined in the beginning of section~\orthsec.
Under the dualities \AFqtogamma\ this identity
implies the relations\foot{To see this we use that $P\cdot(\s\circ\tau)= \big((P\cdot\s)\circ (P\cdot\tau)\big)$ if $P$ has no common letters with $\s$ and
$\tau$ and $\t(\s\circ\tau)=\t(\tau)\circ\t(\s)$ to show that the theorem 4.2 of \garsiareut\ implies $(1\cdot \cI_p)\circ (1\cdot
E^\t_\mu) = 0$ for $\l(p)\neq\mu$ leading to \AYMsymt\ since $\AFq$ corresponds to a sum of
$\t(E_\mu)$ with
partitions with three parts $k(\mu)=3$ and $\AYM$ to $\t(E_\mu)$ with
$k(\mu)=1$.}
\eqn\AYMsymt{
\AYM\big(\g_{1|P_1, \ldots,P_k}\big) = 0\,,\quad k\neq1\,,\qquad
\AFq\big(\g_{1|P_1, \ldots,P_k}\big) = 0\,,\quad k\neq3\,.
}

\newsubsec\Cgamsec BRST-invariant permutations and BRST-invariant superfields

The linearity condition $M_n = \phi\big(A^{\rm string}(P_n)\big)$
implies that BRST-invariant permutations in $P_n$ are mapped to
kinematics in $M_n$ leading to a
correspondence between the BRST-invariant permutations \BRSTdef\ and a series of
motivic MZV corrections from the motivic string disk amplitude in the $f$-alphabet
\eqn\corresp{
\g_{1|P_1,P_2, \ldots,P_k} \leftrightarrow \phi\big(\Astring(\g_{1|P_1,P_2, \ldots,P_k})\big)\,.
}
whose precise content follows from Table 1 and \AstringP.
For example,
\eqnn\idsI
$$\eqalignno{
\g_{1|234} \leftrightarrow & \AYM(1,2,3,4) + \z_3\ap^3s_{12}s_{23}(s_{12}+s_{23})\AYM(1,2,3,4) + \cdots &\idsI\cr
\g_{1|2,3,4} \leftrightarrow &{}-6 \z_2 \ap^2 s_{12}s_{23}\AYM(1,2,3,4) \cr
&{} - 3\z_2^2\ap^4 (s_{12}^3s_{23}+s_{12}^2s_{23}^2 + s_{12}s_{23}^3)\AYM(1,2,3,4) + \cdots \cr
}$$
with similar expansions at higher points.
The data presented in Table 1 and the discussion in section~\AFqsec\ suggest that the
BRST-invariant permutations can be associated to a series of higher-mass BRST-invariant superfields by
defining
\eqn\gamseries{
A_{\z_2^m \z_M}(\g_{1|P_1, \ldots,P_k}) = k! C^{\z_2^m\z_M}_{1|P_1, \ldots,P_k}\,.
}
The motivic MZV amplitudes then follow from
\eqn\apamps{
A_{\z_2^m\z_M}(1,2,3, \ldots,n) = \sum_{k=1}^{2m+1}\sum_{23 \ldots n} C^{\z_2^m\z_M}_{1|P_1,
\ldots,P_k},
}
where $\sum_{23 \ldots n}$ is a shorthand notation for the deconcatenations of $P_1P_2 \ldots P_k = 23 \ldots n$.
For example,
\eqn\Afzf{
A_{\z_2^2}(1,2,3,4)= C^{\z_2^2}_{1|234} + C^{\z_2^2}_{1|2,3,4} = - {2\over5}(
s_{12}s_{23}^3 + {1\over4}s_{12}^2s_{23}^2 + s_{12}^3s_{23})\AYM(1,2,3,4)\,,
}
where the BRST invariants $C^{\z_2^2}_{1|234}=1/10(s_{12}s_{23}^3 + 4 s_{12}^2s_{23}^2 +
s_{12}^3s_{23})\AYM(1,2,3,4)$ and
$C^{\z_2^2}_{1|2,3,4}=-1/2(s_{12}s_{23}^3 + s_{12}^2s_{23}^2 + s_{12}^3s_{23})\AYM(1,2,3,4)$ were
used. The superfield representation of the above BRST invariants is known only for the two
simplest cases, $C_{1|P} = M_1 E_P$ \nptMethod\ and
$C^{\z_2}_{1|P_1,P_2,P_3}$ \refs{\EOMbbs,\partI}.
Note that $C_{1|P_1} = \AYM(1,P_1)$ while the S-map algorithm of \EOMbbs\ gives
rise to a purely combinatorial translation between $C^{\z_2}_{1|P_1,P_2,P_3}$ and sums of
$s_{ij}^2 \AYM$. It remains to be seen whether there exists a general algorithm
to rewrite $C^{\z_2^m \z_M}$ in terms of SYM amplitudes.

\newsubsec\CAsFq BRST invariants from $\AFq$

Another consequence of the duality \Ctogamma\ is that
the representation of $\AFq$ in terms of
$C_{1|P_1,P_2,P_3}$ given in \Afq\ is invertible, as argued indirectly in \oneloopbb.
This follows from Theorem 4.2 of \garsiareut, $E_\mu \circ I_p = I_p$ if $\l(p) = \mu$
where $\l(p)$ is the shape of the composition $p$ and $E_\mu$
is defined in \Elambdadef.
This implies
$(1\cdot \cI_p)\circ (1\cdot \t(E_\mu)) = (1\cdot \cI_p)$ for $\l(p)=\mu$ or, using
the function interpretation of the right action $\s\circ F:=F(\s)$ with
a partition with three parts $k(\mu)=3$
\eqn\projCt{
\AFq\big(\g_{1|P_1,P_2,P_3}\big) = 6C_{1|P_1,P_2,P_3}\,,\quad \len{P_i}=p_i\,,\quad P_1P_2P_3 = 23 \ldots n\,,
}
where we used the identifications \gammacIp, \gamieq\ and duality \AFqtogamma\ on the left-hand side
and the duality \Ctogamma\ on the right-hand side.
For example, from $\gamma_{1|23,4,5}$ of \fivegammaEx\ we get
\eqnn\Ctooproj
$$\eqalignno{
6C_{1|23,4,5} &=
        \AFqf(1,2,3,4,5)
       + \AFqf(1,2,3,5,4)
       + \AFqf(1,2,4,3,5)
       + \AFqf(1,2,4,5,3)
       + \AFqf(1,2,5,3,4)
       + \AFqf(1,2,5,4,3)&\Ctooproj\cr&
       - \AFqf(1,3,2,4,5)
       - \AFqf(1,3,2,5,4)
       - \AFqf(1,3,4,2,5)
       - \AFqf(1,3,5,2,4)
       + \AFqf(1,4,2,3,5)
       - \AFqf(1,4,3,2,5)\,,
}$$
where we used the parity relation \symmetries\ in the RHS.

\newsubsec\gammaCapp The superfield expansion of $C_{1|P,Q,R}$ from BRST-invariant permutations

The so-called BRST invariants $C_{1|P,Q,R}$ of the pure
spinor formalism \psf\ play an important role in the mapping between BRST-invariant permutations
and kinematics, see section~\Cgamsec. They were firstly derived at low multiplicities in \oneloopbb\ and were
subsequently studied in different contexts and given general recursive algorithms,
see \refs{\EOMbbs,\partI,\oneloopI}.
Their superfield expansions in terms of Berends-Giele currents follow from
\eqn\algoC{
C_{i|P,Q,R} = M_i M_{P,Q,R} + M_i\cdot \big[C_{p_1|p_2 \ldots p_\len{P},Q,R} - C_{p_\len{P}|p_1 \ldots p_{\len{P}-1},Q,R} +
(P\leftrightarrow Q,R)\big]
}
starting from $C_{i|j,k,l}=M_i M_{j,k,l}$ with the dot representing concatenation, $M_i\cdot M_A := M_{iA}$.
For example, the first few expansions are given by
\eqnn\Cscex
$$\eqalignno{
C_{1|2,3,4} &= M_1 M_{2,3,4}\,, &\Cscex \cr
C_{1|23,4,5} &=M_1 M_{23,4,5} + M_{12} M_{3,4,5} - M_{13} M_{2,4,5}\,,  \cr
C_{1|234,5,6} &= M_1 M_{234,5,6} + M_{12}M_{34,5,6} + M_{123}M_{4,5,6} - M_{124}M_{3,5,6}\cr
&\quad{}- M_{14}M_{23,5,6} - M_{142}M_{3,5,6} + M_{143}M_{2,5,6}\,, \cr
C_{1|23,45,6}
&= M_1 M_{23,45,6} + M_{12}M_{45,3,6} - M_{13}M_{45,2,6} + M_{14}M_{23,5,6} - M_{15}M_{23,4,6}\cr
&\quad{}+M_{124}M_{3,5,6} - M_{134}M_{2,5,6}+ M_{142}M_{3,5,6} - M_{152}M_{3,4,6}\cr
&\quad{}-M_{125}M_{3,4,6} + M_{135}M_{2,4,6} - M_{143}M_{2,5,6} + M_{153}M_{2,4,6}\,.
}$$
These terms can be extracted from the permutations of the BRST-invariant permutations
$\g_{1|P,Q,R}$
of the inverse descent algebra as follows:
\medskip
\item{1.} Sum over the cyclic permutations of all permutations in $\g_{1|P,Q,R}$:
\eqn\sumc{
W_\s\to W_\s + {\rm cyclic}(\s)
}
\item{2.} Decompose $W_\s$ into all possible four-word deconcatenations:
\eqn\decf{
W_\s = \sum_{XYZW=\s}W_X.W_Y.W_Z.W_W
}
\item{3.} Move label $1$ to the front by repeatedly commuting $W_{C}.W_{A1B} = W_{A1B}.W_C$ if necessary
and write the result in terms of Berends-Giele superfields:
\eqn\toBG{
W_{A1B}.W_C.W_D.W_E := {1\over4!}M_{A1B}M_{C,D,E}
}
\medskip
\noindent The resulting expressions
have been explicitly checked\foot{The shuffle symmetry $AiB=(-1)^\len{A}i \tilde A \shuffle B$ \BGschocker\
is needed to rewrite words in a Lyndon basis.} for all topologies of BRST invariants
up to eight points. In addition, using the descent duality \Ctogamma\ one may also derive the change of
basis identities for $C_{i\neq 1| \ldots} = \sum C_{1| \ldots}$ from \refs{\partI,\oneloopI} by choosing a different
label to be singled-out in the color-dressed permutation \cdressedP\ \rugpriv.

\newnewsec\concsec Conclusion

In this paper we investigated the combinatorial properties of the permutations appearing in the
{\it color-dressed permutations} \cdressedP\ using the tools from the descent algebra.
In particular, we defined BRST-invariant permutations, found a closed formula, and related them
to orthogonal idempotents
which sum to the identity permutation \refs{\PBWReutenauer,\garsia,\garsiareut}.

We then considered the color-dressed motivic string disk amplitudes of \motivic\
within this framework. This led to the discovery of the relations \AYMlike\
obeyed by the motivic MZV amplitudes, dubbed the {\it descent algebra decomposition of KK-like relations}.
The basis dimensions of linearly independent motivic MZV amplitudes
are given by sums of Stirling cycle numbers.
These claims have been explicitly checked
using various data points in string theory up to $n=8$ and $\ap^7$.

Inspired by \oneloopbb, we proposed a correspondence
between the permutations from the (inverse) descent algebra and kinematics from the motivic string disk
amplitudes in terms of higher-mass BRST invariants. In the particular case of $\ap^2$,
we exploited a
theorem from the mathematics literature on descent algebra to prove certain claims in \oneloopbb\
and to systematically express BRST invariants as linear combinations of
$A_{\z_2}$ corrections
to motivic disk amplitudes in \projCt.

And finally, we found an algorithm to extract the superfield content of the BRST invariants
in the pure spinor formalism from
the BRST-invariant permutations in the inverse descent algebra.
It will be interesting to obtain superfield realizations of the
higher-mass BRST
invariants defined in section~\Cgamsec. They can probably be extracted from a perturbiner series of amplitudes at the
appropriate mass level. For instance, the superfields in $C^{\rm SYM}_{1|P}=M_1E_P$ are related to the
series $\Tr(\bV\bV\bV)$ \refs{\BGBCJ,\SYM}. The superfields in $C^{\z_2}_{1|P_1,P_2,P_3}$ are related to the
series $\Tr(\bV_1(\l\g^m \bW)(\l\g^n \bW)\bF_{mn})$, while the superfields in $C^{\z_3}_{1|P}$
should follow from $\Tr\big((\l\g^{mnpqr}\l)(\l\g^s\bW)\bF_{mn}\bF_{pq}\bF_{rs}\big)$. It would also be
interesting to find combinatorial algorithms to directly translate the higher-mass BRST invariants
$C^{\z_2^m\z_M}_{1|P_1, \ldots,P_k}$ into linear
combinations of super-Yang--Mills amplitudes and powers of Mandelstam invariants, generalizing
the S-map algorithm of \EOMbbs\ for $C^{\z_2}_{1|P_1,P_2,P_3}$.
This would give rise to a combinatorial description of the $P_n$ and $M_n$ matrices of \motivic.

\bigskip
\noindent{\bf Acknowledgements:}
I thank Ruggero Bandiera for discussions during
an early stage of this project. I also thank
Oliver Schlotterer for sharing notes containing intriguing observations about
the permutations in the color-dressed amplitude that served as the
motivation for this paper, for discussions, for collaboration on related topics, and
for comments on the draft (including suggesting a v3 to be uploaded).
CRM is supported by a University Research Fellowship from the Royal Society.

\appendix{A}{The Solomon descent algebra}
\applab\SolomonDSsec

\noindent We review the salient features of the Solomon descent algebra
\refs{\mackey,\garsiareut,\PBWReutenauer,\schockerSolomon,\garsiaremmel,\thibon,\Reutenauer}.
In particular, we discuss different bases and highlight
the orthogonal idempotents discovered by Reutenauer, as they will be related to $\ap$ corrections
to string amplitudes.

\newsubsecstar\descsec Descent classes and the Solomon descent algebra

The {\it descent set} $D(\s)$ and the and the {\it descent number\/} $d_\s$
of a permutation $\sigma=\sigma_1\sigma_2 \ldots\sigma_n$ in $S_n$
are defined by
\eqn\Ddef{
D(\sigma) = \{i\in \{1,2, \ldots,n-1\}\mid \si_i> \si_{i+1}\}\,,\qquad d_\s = \#\big(D(\s)\big)\,.
}
For example, the permutation $\s=546132$ has descent set $D(\s)=\{1,3,5\}$ and descent number $d_\s=3$.
The collection of permutations with a given descent set $S$ is called a {\it descent class},
\eqn\DSdef{
D_S = \sum_{D(\si)=S} \si\,.
}
For example, the permutations in $S_3$ are distributed into four descent classes,
\eqn\Sthree{
D_\emptyset = W_{123},\quad D_{\{1\}} = W_{213}+W_{312},\quad D_{\{2\}} = W_{132}+W_{231},\quad D_{\{1,2\}} = W_{321}\,.
}
In general, the permutations of $S_n$ decompose into $2^{n-1}$ distinct descent classes; all the subsets in
the powerset of $\{1,2, \ldots,n-1\}$ since the last $n$-th position is never a descent.
Solomon showed the remarkable property that descent classes are closed under the right action \rightaction
\eqn\Dalg{
D_S\circ D_T = \sum_{U\subseteq \{1,2, \ldots,n-1\}} c_{S,T,U}D_U
}
where the coefficients $c_{S,T,U}$ are non-negative integers \mackey.
The descent classes therefore form a $2^{n-1}$ dimensional algebra, the so-called {\it Solomon's
descent algebra} ${\cal D}_n$ \refs{\mackey,\garsiareut,\schockerSolomon,\garsiaremmel,\thibon,\Reutenauer}.

As an example of \Dalg, consider the permutations in $S_4$.
Its 24 elements are organized into 8 descent classes as follows
\eqn\twoDs{
\eqalign{
D_{\{\emptyset\}} &= \Wordq(1,2,3,4)\,,\cr
D_{\{1\}} &=
        W_{2134}
       + W_{3124}
       + W_{4123}\,, \cr
D_{\{2\}} &=
        W_{1324}
       + W_{1423}
       + W_{2314}
       + W_{2413}
       + W_{3412}\,,\cr
D_{\{1,3\}} & = \Wordq(2,1,4,3)
       + \Wordq(3,1,4,2)
       + \Wordq(3,2,4,1)
       + \Wordq(4,1,3,2)
       + \Wordq(4,2,3,1)\,,\cr
}\quad\eqalign{
D_{\{1,2\}} &=
 	\Wordq(3,2,1,4)
       + \Wordq(4,2,1,3)
       + \Wordq(4,3,1,2)\,,\cr
D_{\{3\}} &=
	\Wordq(1,2,4,3)
       + \Wordq(1,3,4,2)
       + \Wordq(2,3,4,1)\,,\cr
D_{\{2,3\}} &=
	 \Wordq(1,4,3,2)
       + \Wordq(2,4,3,1)
       + \Wordq(3,4,2,1)\,,\cr
D_{\{1,2,3\}} & = \Wordq(4,3,2,1)\,.
}}
It is straightforward to multiply the permutations among these descent classes
using the right-action of the symmetric group \rightaction. For example,
\eqnn\DoDt
$$\eqalignno{
D_{\{1\}}\circ D_{\{2\}}&=
        W_{1234}
       + W_{1243}
       + W_{1324}
       + W_{1342}
       + W_{1423}
       + W_{1432}
       + W_{2314}&\DoDt\cr
&       + W_{2341}
       + W_{2413}
       + W_{2431}
       + W_{3214}
       + W_{3412}
       + W_{3421}
       + W_{4213}
       + W_{4312}\,\cr
& = D_{\{\emptyset\}}
       + D_{\{1,2\}}
       + D_{\{2\}}
       + D_{\{2,3\}}
       + D_{\{3\}}\,,
}$$
where the last line follows from
the remarkable property \Dalg\ which ensures that the permutations in \DoDt\
are themselves a sum of descent classes.

\newsubsecstar\basissec Bases of the descent algebra

Apart from the
descent classes $D_S$ indexed by descent sets $S$, there are other convenient bases of
the descent algebra \garsiareut.

\newsubsubsec\Bpbasissec Composition basis $B_p$

The composition $p$ of $n$, denoted $p\models n$, is a $k$-tuple of positive integers with
sum $n$,
\eqn\pdef{
p=(p_1,p_2, \ldots,p_k),\quad p_1+p_2+ \cdots + p_k = n.
}
There is a bijection between compositions $p\models n$ and subsets $S$ of $\{1,2, \ldots, n-1\}$
\eqnn\ptoS
\eqnn\Stop
$$\eqalignno{
p = (p_1,p_2, \ldots,p_k) &\mapsto \{p_1,\, p_1+p_2, \ldots,\, p_1 + p_2 + \cdots + p_{k-1}\} := S(p)\,,&\ptoS\cr
S=\{i_1,i_2, \ldots,i_k\} &\mapsto (i_1,i_2-i_1, \ldots, i_k - i_{k-1},n - i_k) :=C_n(S)\,. &\Stop
}$$
Thus the total number of compositions of $n$ is $2^{n-1}$,
the cardinality of the powerset of $\{1,2, \ldots,n-1\}$.
Note that the map $C_n(S)=p$ depends on the order $n$ of the permutation group $S_n$ for
$C_3\big(\{1,2\}\big) = (1,1,1)$ but $C_4\big(\{1,2\}\big) = (1,1,2)$. In particular, $C_n(\emptyset) = (n)$.

The basis $B_p$ is indexed by compositions $p$ rather than subsets and is defined by \garsiareut
\eqn\Bpdef{
B_p = D_{\subseteq S(p)}\,,
}
with $S(p)$ given by \ptoS.
For example, the $D_S$ basis elements \twoDs\ become
\eqn\BpToDex{
\eqalign{
B_{1111} &= D_\emptyset + D_{\{1\}} + D_{\{2\}} + D_{\{3\}} + D_{\{1,2\}}\cr
&+ D_{\{1,3\}} + D_{\{2,3\}} + D_{\{1,2,3\}}\,,\cr
B_{112} &= D_\emptyset + D_{\{1\}} + D_{\{2\}}+ D_{\{1,2\}}\,,\cr
B_{121} &= D_\emptyset + D_{\{1\}} + D_{\{3\}}+ D_{\{1,3\}}\,,\cr
B_{211} &= D_\emptyset + D_{\{2\}} + D_{\{3\}}+ D_{\{2,3\}}\,,
}\quad\eqalign{
B_{13} &= D_\emptyset + D_{\{1\}}\,,\cr
B_{22} &= D_\emptyset + D_{\{2\}}\,,\cr
B_{31} &= D_\emptyset + D_{\{3\}}\,,\cr
B_4 &= D_{\emptyset}\,.
}}
The inverse of \Bpdef\ is
given by Lemma 8.18 in \Reutenauer
\eqn\BpToDS{
D_S = \sum_{T\subseteq S}(-1)^{\len{S}-\len{T}}D_{\subseteq T}\,.
}
For example (in $S_4$),
$D_{\{1,2\}} = B_{112}-B_{13}- B_{22}+B_4$, $D_{\{1\}} = B_{13}- B_4$,
$D_{\{2\}} = B_{22}- B_4$, and $D_\emptyset = B_4$,
from which we verify that $D_\emptyset + D_{\{1\}} + D_{\{2\}}+ D_{\{1,2\}} = B_{112}$.

The permutations within a basis element $B_p$ can be found via \refs{\garsia,\garsiareut}
\eqn\BpToPermutations{
B_{p_1p_2 \ldots p_k} =\theta(X_1\shuffle X_2\shuffle \ldots \shuffle X_k)\,,\quad 12 \ldots n = X_1 \ldots X_k,\quad |X_i|=p_i\,.
}
where the inverse map $\t$ is given by
\eqn\thetamap{
\theta(\sigma) \mapsto \sigma^{-1}\,.
}
For example, if $p=(1,1,2)$ then $X_1=1$, $X_2=2$ and
$X_3=34$ and we get
\eqnn\Bexamp
$$\eqalignno{
B_{112}=\theta(1\shuffle 2\shuffle 34)
&=	 W_{1234}
       + W_{1324}
       + W_{1423}
       + W_{2134}
       + W_{2314}
       + W_{2413}&\Bexamp\cr
&       + W_{3124}
       + W_{3214}
       + W_{3412}
       + W_{4123}
       + W_{4213}
       + W_{4312}\,.
}$$

\newsubsubsec\multsec Multiplication table for $B_p\circ B_q$

There is a closed formula for the multiplication of $B_p\circ B_q$ \refs{\garsiaremmel,\garsiareut,\atkinson}.
Let $M$ be a matrix with non-negative integer entries $m_{ij}$ whose
{\it row sum} $r(M)$ and {\it column sum} $c(M)$ are vectors defined by
\eqn\rcsum{
r(M)_i := \sum_j m_{ij}\,,\qquad c(M)_j := \sum_i m_{ij}\,.
}
Then
\eqn\BpBq{
B_p\circ B_q = \sum_{c(M) = p\atop r(M) = q} B_{{\rm co}(M)}
}
where ${\rm co}(M)$ denotes the composition obtained by reading the matrix $M$ row by row from top to bottom while excluding
the zero entries $m_{ij}=0$.
This product is associative and
$B_{n}$ is a multiplicative identity for compositions of $n$ \atkinson.

For example, let us recover the result \DoDt\ for $D_{\{1\}}\circ D_{\{2\}}$ using the above multiplication table
\BpBq\ in $S_4$. Given that $D_{\{1\}}=B_{13}-B_{4}$ and $D_{\{2\}}=B_{22}-B_{4}$, the only non-trivial
product we need is $B_{13}\circ B_{22}$ since
$B_4$ is the identity for compositions of $n=4$.
The set of integer matrices $M$ with $c(M)=(1,3)$ and $r(M)=(2,2)$ is given by
\eqn\Zot{
\pmatrix{ 1 & 1\cr 0 & 2}\,,\qquad
\pmatrix{ 0 & 2\cr 1 & 1}\,.
}
Thus
$B_{13}\circ B_{22} = B_{112} + B_{211}$
and $\Du(1)\circ\Du(2) = (B_{13}-B_4)\circ(B_{22}-B_4)$ implies
\eqn\Dse{
\Du(1)\circ\Du(2) = B_{112}+B_{211}-B_{13}-B_{22}+B_4 = D_{\emptyset}+D_{\{1,2\}}+D_{\{2\}}+D_{\{2,3\}}+D_{\{3\}}
}
where we used the conversions \BpToDex.

\newsubsubsec\eulidsec The Eulerian idempotent

The Eulerian (or Solomon) idempotent is defined by \refs{\solomon,\PBWReutenauer,\garsia,\Loday} (see also \schatz)
\eqn\eulsi{
E_n = \sum_{\s\in S_n}\kappa_\s\s,\qquad \kappa_\s = {(-1)^{d_\s} \over|\s|{|\s|-1\choose d_\s}}
}
where $d_\s$ denotes the descent number \Ddef\ of the permutation $\s$. For example,
\eqn\Euex{
E_2 = \half\big(W_{12}-W_{21}\big),\quad
E_3 = {1\over3} \Word(1,2,3)
       - {1\over6} \Word(1,3,2)
       - {1\over6} \Word(2,1,3)
       - {1\over6} \Word(2,3,1)
       - {1\over6} \Word(3,1,2)
       + {1\over3} \Word(3,2,1)\,.
}
Apart from being an idempotent satisfying $E_n\circ E_n = E_n$,
the definition \eulsi\ is also a Lie polynomial \PBWReutenauer.
Therefore its coefficients $\kappa_\s$ must satisfy the shuffle symmetry \Ree
\eqn\kapshu{
\kappa_{R\shuffle S} = 0,\quad R,S\neq\emptyset\,.
}
As usual, the definition \eulsi\ in terms of the fixed alphabet ${\Bbb N}$ in $S_n$ can be turned into
a {\it function} of an arbitrary word $P$ by the right action \rightaction\ of the
symmetric group \refs{\Reutenauer,\malvenutoreutenauer},
\eqn\eulfunc{
E(P) = E^P := P\circ E_n, \quad |P|=n\,.
}
For example,
$E(i,j,k) = ijk\circ E_3 = {1\over3} \Word(i,j,k)
       - {1\over6} \Word(i,k,j)
       - {1\over6} \Word(j,i,k)
       - {1\over6} \Word(j,k,i)
       - {1\over6} \Word(k,i,j)
       + {1\over3} \Word(k,j,i)$.

\newsubsubsec\Idbasis The idempotent basis $I_p$

The idempotent basis $I_p$ of the descent algebra ${\cal D}_n$ satisfying $I_p\circ I_p = I_p$ was introduced in
\garsiareut\ and it is indexed by the compositions of $n$
\eqn\Ipdef{
I_{p_1p_2 \ldots p_k}(P) = \sum_{X_1, \ldots,X_k\atop
\len{X_i}=p_i} \langle P,X_1\shuffle X_2\shuffle \ldots \shuffle X_k\rangle
E^{X_1}E^{X_2} \ldots E^{X_k}\,,
}
where the sum is constrained by the length of $X_i$ being equal to the corresponding $p_i$ in the composition $p$
and $E^{X_i}$ denote the Eulerian idempotent function \eulfunc.
For example, with canonical $P=12 \ldots n$ we have
\eqnn\Ipex
$$\eqalignno{
I_{11} &=W_{12}+W_{21}\,,\qquad
I_2=\half(W_{12}-W_{21})\,,&\Ipex\cr
I_{111} & = W_{123} + W_{132} + W_{213} + W_{231} + W_{312} + W_{321}\,,\cr
I_{21} &= {1\over2} W_{123}
       + {1\over2} W_{132}
       - {1\over2} W_{213}
       + {1\over2} W_{231}
       - {1\over2} W_{312}
       - {1\over2} W_{321}\,,\cr
I_{12} &= {1\over2} W_{123}
       - {1\over2} W_{132}
       + {1\over2} W_{213}
       - {1\over2} W_{231}
       + {1\over2} W_{312}
       - {1\over2} W_{321}\,,\cr
I_3 &=  {1\over3} W_{123}
       - {1\over6} W_{132}
       - {1\over6} W_{213}
       - {1\over6} W_{231}
       - {1\over6} W_{312}
       + {1\over3} W_{321}\,.
}$$

\newsubsubsec\IptoBpsec $I_p$ to $B_p$

The idempotent basis elements $I_p$ for $p=p_1p_2 \ldots p_k$ can be expanded in terms of compositions $B_q$ using an
algorithm discussed in \garsiareut. First one defines {\it moments} $e_m$ as a polynomial in non-commuting variables
$t_i$ for $i=1,2, \ldots$ from the generating series
\eqn\moment{
\sum x^n e_n = \log(1+\sum t_i x^i)
}
where $x$ is a commuting parameter. For example, from \moment\ it follows that
\eqnn\enex
$$\displaylines{
e_1 = t_1,\quad e_2 = t_2 - \half t_1^2,\quad
e_3 = t_3 - \half(t_1t_2+t_2t_1)+{1\over3}t_1^3 \hfil\enex\hfilneg\cr
e_4 = - {1\over4} t_1^4
          + {1\over3} t_1^2 t_2
          + {1\over3} t_1t_2t_1
          - {1\over2} t_1t_3
          + {1\over3} t_2 t_1^2
          - {1\over2} t_2^2
          - {1\over2} t_3 t_1
          + t_4
}$$
Then to convert the $I_p$ basis elements to the composition basis $B_q$ one uses \garsiareut
\eqn\IpToBp{
I_p = \delta(e_{p_1}e_{p_2} \ldots e_{p_k}),\hbox{ with }\d(t_{i_1}t_{i_2} \ldots t_{i_k}) := B_{i_1i_2 \ldots i_k}.
}
For example,
\eqn\Ifour{
I_4 = - {1\over4} B_{1111}
       + {1\over3} B_{112}
       + {1\over3} B_{121}
       - {1\over2} B_{13}
       + {1\over3} B_{211}
       - {1\over2} B_{22}
       - {1\over2} B_{31}
       + B_{4}\,.
}

\newsubsecstar\orthsec Reutenauer orthogonal idempotents

A partition $\l$ of $n$, denoted $\l\vdash n$, is a $k$-tuple of positive integers with sum~$n$ 
satisfying $\l_1\ge \l_2\ge \ldots \ge \l_k$.
If $p\models n$ is a composition of $n$, the {\it shape} $\lambda(p)$ of $p$
is the partition of $n$ obtained by rearranging the parts of $p$ in decreasing order.
Also, $k(p)$ is the {\it number of parts} of the composition $p$.
For example,
$p=(2,3,1,2)$ implies
$\l(p) = 3221$ and $k(p)=4$.

Given a partition $\l=(\l_1,\l_2,
\ldots,\l_k)$ into $k$ parts, theorem 3.1 of \garsiareut\ shows that
\eqn\Elambdadef{
E_\l := {1\over k!}\sum_{\l(p) = \l}I_p\,,\qquad \sum_{\l\vdash n}E_\lambda = W_{12 \ldots n}\,.
}
Note that when the partition $\l$ of $n$ has only one part, $E_\l=I_n$ coincides with the Eulerian
idempotent $E_n$ \eulsi, so this notation is not ambiguous.
For example, $E_1=I_1$ and
\eqn\Eexs{
\eqalign{
E_2 &= I_2\,,\cr
E_{11}&=\half I_{11}\,,
}\qquad
\eqalign{
E_{3} &=I_{3}\,,\cr
E_{21} &={1\over2} \big(I_{12}+I_{21}\big)\,,
}\qquad\eqalign{
E_{111} &={1\over3!} I_{111}\,,\cr
E_{211} &={1\over3!} \big(I_{112} + I_{121} + I_{211}\big)\,.
}}
one can readily verify $E_3 + E_{21} + E_{111} = W_{123}$
using the expansions listed in the appendix~\higherap.

The Reutenauer idempotents $E^{(m)}$ are defined in the alphabet
$\{1,2, \ldots\}$ as the sum over all
permutations of $E_\lambda$ from \Elambdadef\ such that $\l$ is a partition of $n$ with $m$
parts, i.e.,
\eqn\Ereutdef{
E^{(m)} = \sum_{\l\vdash n\atop k(\l)=m} E_\l
}
For example,
\eqn\Etwos{
\eqalign{ n&=2\cr n&=3\cr n&=4}\qquad\eqalign{
E^{(1)} &= E_2,\quad E^{(2)}= E_{11}\cr
E^{(1)} &= E_3,\quad E^{(2)}= E_{21},\quad E^{(3)}=E_{111}\cr
E^{(1)} &= E_4,\quad E^{(2)}= E_{31}+E_{22},\quad E^{(3)}=E_{211},\quad E^{(4)}=E_{1111}
}}
It was shown in \refs{\garsiareut,\PBWReutenauer} that \Ereutdef\ are orthogonal
idempotents which sum to the identity permutation
\eqn\reutorth{
\sum_{i=1}^n E^{(i)} = W_{123 \ldots n}\,,\qquad
E^{(i)}\circ E^{(j)} =
\cases{
E^{(i)} & if $i=j$;\cr
0 & otherwise.}
}
An alternative definition of the Reutenauer
idempotents in terms of a generating function can be found in \Reutenauer.

\appendix{B}{The inverse of the idempotent basis}
\applab\cIpap

\noindent In this appendix we will prove \proptIp, that is:

\proclaim Proposition. The inverse of the idempotent basis \cIpdef\ satisfies
\eqn\proptIpapp{
\cI_{p_1p_2 \ldots p_k}(P_1,P_2, \ldots,P_k)
= \cE(P_1)\shuffle\cE(P_2)\shuffle \ldots\shuffle \cE(P_k)\,,
\quad\len{P_i}=p_i
}
where $P=P_1 \ldots P_k$ is the factorization of $P$ with $P_i$ of length $p_i$.
\par
\noindent{\it Proof.} The proof will be based on the following observations collected from \Reutenauer, which
should be consulted for more details as the equation numbers below refer to it.
First, the adjoint of an arbitrary function $F(P)=P\circ F$ of a word $P$ is given
by $\t(F)(P) = P\circ\t(F)$, see (3.3.5).
Second, the adjoint of $F_{p_1}\star F_{p_2} \ldots \star F_{p_k}$
is $\t(F_{p_1})\star' \t(F_{p_2}) \ldots \star' \t(F_{p_k})$ where $\star$ and $\star'$ are the convolution
operators  defined in (1.5.7) and (1.5.8) and
$\t(F_j)$ is the adjoint of $F_j$ when viewed as a function by the right-action \rightaction, see proof of Lemma 3.13.
Third, for permutations $F_{p_i}$ of length $p_i$
one can show (by adapting the proof of Lemma 3.13)
\eqn\essen{
\big(F_{p_1}\star' \ldots \star'F_{p_k}\big)\!(P)= F_{p_1}\!(P_1)\shuffle \ldots \shuffle F_{p_k}\!(P_k)
}
where the functions are defined via a right action as $F_{p_i}(P_i):=P_i\circ F_{p_i}$.
The proof of \proptIpapp\ then follows from the observation by (1.5.4) and (1.5.7) that the idempotent basis $I_p$ \Ipdef\ can
be rewritten as a convolution
$I_{p_1 \ldots p_k}(P) = \big(E_{p_1}\star \ldots \star E_{p_k}\big)\big(P\big)$
where $E_p$ is the Eulerian idempotent \eulsi. Therefore its adjoint $\t(I_{p_1 \ldots p_k})(P)$
is given by
\eqnn\adjIp
$$\eqalignno{
\t(I_{p_1 \ldots p_k})(P) &= \t(E_{p_1})(P_1)\star' \ldots \star'\t(E_{p_k})(P_k),\quad
P=P_1 \ldots P_k,\quad\len{P_i}=p_i \cr
&= \cE(P_1)\shuffle \ldots \shuffle \cE(P_k)&\adjIp
}$$
where we used \essen\ and $\cE(P_i)=\t(E_{p_i})(P_i)$. \QED

As a multiplicity-four example of the inverse idempotent basis of \cIpdef\ we have
\eqnn\cIptwotwo
$$\eqalignno{
\cI_{22}(12,34) &=
        {1\over4} \Wordq(1,2,3,4)
       - {1\over4} \Wordq(1,2,4,3)
       + {1\over4} \Wordq(1,3,2,4)
       + {1\over4} \Wordq(1,3,4,2)
       - {1\over4} \Wordq(1,4,2,3)
       - {1\over4} \Wordq(1,4,3,2)&\cIptwotwo\cr&
       - {1\over4} \Wordq(2,1,3,4)
       + {1\over4} \Wordq(2,1,4,3)
       - {1\over4} \Wordq(2,3,1,4)
       - {1\over4} \Wordq(2,3,4,1)
       + {1\over4} \Wordq(2,4,1,3)
       + {1\over4} \Wordq(2,4,3,1)\cr&
       + {1\over4} \Wordq(3,1,2,4)
       + {1\over4} \Wordq(3,1,4,2)
       - {1\over4} \Wordq(3,2,1,4)
       - {1\over4} \Wordq(3,2,4,1)
       + {1\over4} \Wordq(3,4,1,2)
       - {1\over4} \Wordq(3,4,2,1)\cr&
       - {1\over4} \Wordq(4,1,2,3)
       - {1\over4} \Wordq(4,1,3,2)
       + {1\over4} \Wordq(4,2,1,3)
       + {1\over4} \Wordq(4,2,3,1)
       - {1\over4} \Wordq(4,3,1,2)
       + {1\over4} \Wordq(4,3,2,1)\,.
}$$

\appendix{C}{Explicit permutations at low multiplicities}
\applab\higherap

\noindent The multiplicity-five BRST-invariant permutations (defined in \BRSTdef) are given by
\eqnn\fivegammaEx
$$\eqalignno{
\gamma_{1|2,3,4,5} &=W_{1(2\shuffle3\shuffle4\shuffle5)}&\fivegammaEx\cr
\gamma_{1|23,4,5} &=
	    {1\over2} W_{12345}
          + {1\over2} W_{12354}
          + {1\over2} W_{12435}
          + {1\over2} W_{12453}
          + {1\over2} W_{12534}
          + {1\over2} W_{12543}\cr&
          - {1\over2} W_{13245}
          - {1\over2} W_{13254}
          - {1\over2} W_{13425}
          - {1\over2} W_{13452}
          - {1\over2} W_{13524}
          - {1\over2} W_{13542}\cr&
          + {1\over2} W_{14235}
          + {1\over2} W_{14253}
          - {1\over2} W_{14325}
          - {1\over2} W_{14352}
          + {1\over2} W_{14523}
          - {1\over2} W_{14532}\cr&
          + {1\over2} W_{15234}
          + {1\over2} W_{15243}
          - {1\over2} W_{15324}
          - {1\over2} W_{15342}
          + {1\over2} W_{15423}
          - {1\over2} W_{15432}
\cr
\gamma_{1|234,5} &=
            {1\over3} W_{12345}
          + {1\over3} W_{12354}
          - {1\over6} W_{12435}
          - {1\over6} W_{12453}
          + {1\over3} W_{12534}
          - {1\over6} W_{12543}\cr&
          - {1\over6} W_{13245}
          - {1\over6} W_{13254}
          - {1\over6} W_{13425}
          - {1\over6} W_{13452}
          - {1\over6} W_{13524}
          - {1\over6} W_{13542}\cr&
          - {1\over6} W_{14235}
          - {1\over6} W_{14253}
          + {1\over3} W_{14325}
          + {1\over3} W_{14352}
          - {1\over6} W_{14523}
          + {1\over3} W_{14532}\cr&
          + {1\over3} W_{15234}
          - {1\over6} W_{15243}
          - {1\over6} W_{15324}
          - {1\over6} W_{15342}
          - {1\over6} W_{15423}
          + {1\over3} W_{15432}
          \cr
\gamma_{1|23,45} &=
           {1\over4} W_{12345}
          - {1\over4} W_{12354}
          + {1\over4} W_{12435}
          + {1\over4} W_{12453}
          - {1\over4} W_{12534}
          - {1\over4} W_{12543}\cr&
          - {1\over4} W_{13245}
          + {1\over4} W_{13254}
          - {1\over4} W_{13425}
          - {1\over4} W_{13452}
          + {1\over4} W_{13524}
          + {1\over4} W_{13542}\cr&
          + {1\over4} W_{14235}
          + {1\over4} W_{14253}
          - {1\over4} W_{14325}
          - {1\over4} W_{14352}
          + {1\over4} W_{14523}
          - {1\over4} W_{14532}\cr&
          - {1\over4} W_{15234}
          - {1\over4} W_{15243}
          + {1\over4} W_{15324}
          + {1\over4} W_{15342}
          - {1\over4} W_{15423}
          + {1\over4} W_{15432}\cr
\gamma_{1|2345} &=
	   {1\over4} W_{12345}
          - {1\over12} W_{12354}
          - {1\over12} W_{12435}
          - {1\over12} W_{12453}
          - {1\over12} W_{12534}
          + {1\over12} W_{12543}\cr&
          - {1\over12} W_{13245}
          + {1\over12} W_{13254}
          - {1\over12} W_{13425}
          - {1\over12} W_{13452}
          + {1\over12} W_{13524}
          + {1\over12} W_{13542}\cr&
          - {1\over12} W_{14235}
          - {1\over12} W_{14253}
          + {1\over12} W_{14325}
          + {1\over12} W_{14352}
          - {1\over12} W_{14523}
          + {1\over12} W_{14532}\cr&
          - {1\over12} W_{15234}
          + {1\over12} W_{15243}
          + {1\over12} W_{15324}
          + {1\over12} W_{15342}
          + {1\over12} W_{15423}
          - {1\over4} W_{15432}
\cr
}$$
According to the deconcatenation \gamieq\ these BRST-invariant permutations
give rise to the following orthogonal idempotents:
$$\eqalignno{
\g^{(1)}_{12345} &=
            {1\over4} W_{12345}
          - {1\over12} W_{12354}
          - {1\over12} W_{12435}
          - {1\over12} W_{12453}
          - {1\over12} W_{12534}
          + {1\over12} W_{12543}\cr&
          - {1\over12} W_{13245}
          + {1\over12} W_{13254}
          - {1\over12} W_{13425}
          - {1\over12} W_{13452}
          + {1\over12} W_{13524}
          + {1\over12} W_{13542}\cr&
          - {1\over12} W_{14235}
          - {1\over12} W_{14253}
          + {1\over12} W_{14325}
          + {1\over12} W_{14352}
          - {1\over12} W_{14523}
          + {1\over12} W_{14532}\cr&
          - {1\over12} W_{15234}
          + {1\over12} W_{15243}
          + {1\over12} W_{15324}
          + {1\over12} W_{15342}
          + {1\over12} W_{15423}
          - {1\over4} W_{15432}\cr
   \g^{(2)}_{12345} &=
            {11\over24} W_{12345}
          - {1\over24} W_{12354}
          - {1\over24} W_{12435}
          - {1\over24} W_{12453}
          - {1\over24} W_{12534}
          - {1\over24} W_{12543}\cr&
          - {1\over24} W_{13245}
          - {1\over24} W_{13254}
          - {1\over24} W_{13425}
          - {1\over24} W_{13452}
          - {1\over24} W_{13524}
          - {1\over24} W_{13542}\cr&
          - {1\over24} W_{14235}
          - {1\over24} W_{14253}
          - {1\over24} W_{14325}
          - {1\over24} W_{14352}
          - {1\over24} W_{14523}
          - {1\over24} W_{14532}\cr&
          - {1\over24} W_{15234}
          - {1\over24} W_{15243}
          - {1\over24} W_{15324}
          - {1\over24} W_{15342}
          - {1\over24} W_{15423}
          + {11\over24} W_{15432}\cr
   \g^{(3)}_{12345} &=
            {1\over4} W_{12345}
          + {1\over12} W_{12354}
          + {1\over12} W_{12435}
          + {1\over12} W_{12453}
          + {1\over12} W_{12534}
          - {1\over12} W_{12543}\cr&
          + {1\over12} W_{13245}
          - {1\over12} W_{13254}
          + {1\over12} W_{13425}
          + {1\over12} W_{13452}
          - {1\over12} W_{13524}
          - {1\over12} W_{13542}\cr&
          + {1\over12} W_{14235}
          + {1\over12} W_{14253}
          - {1\over12} W_{14325}
          - {1\over12} W_{14352}
          + {1\over12} W_{14523}
          - {1\over12} W_{14532}\cr&
          + {1\over12} W_{15234}
          - {1\over12} W_{15243}
          - {1\over12} W_{15324}
          - {1\over12} W_{15342}
          - {1\over12} W_{15423}
          - {1\over4} W_{15432}\cr
   \g^{(4)}_{12345} &=
	  {1\over24}W_{1(2\shuffle 3 \shuffle 4 \shuffle 5)}
}$$
Here we list the first few expansions of $E_\lambda$ defined in \Elambdadef:
\eqnn\firstElambdas
$$\eqalignno{
E_2 &=
        {1\over 2} \Wordd(1,2)
       - {1\over 2} \Wordd(2,1)\,,\quad\qquad
E_{11} =
        {1\over 2} \Wordd(1,2)
       + {1\over 2} \Wordd(2,1) &\firstElambdas\cr
  E_3 &=
        {1\over 3} \Word(1,2,3)
       - {1\over 6} \Word(1,3,2)
       - {1\over 6} \Word(2,1,3)
       - {1\over 6} \Word(2,3,1)
       - {1\over 6} \Word(3,1,2)
       + {1\over 3} \Word(3,2,1)\cr
  E_{21} &=
        {1\over 2} \Word(1,2,3)
       - {1\over 2} \Word(3,2,1)\,,\quad\qquad
E_{111} =
        {1\over 6} \Word(1,2,3) + {\rm perm}(1,2,3)\cr
E_{211} &=
        {1\over 4} \Wordq(1,2,3,4)
       + {1\over 12} \Wordq(1,2,4,3)
       + {1\over 12} \Wordq(1,3,2,4)
       + {1\over 12} \Wordq(1,3,4,2)
       + {1\over 12} \Wordq(1,4,2,3)
       - {1\over 12} \Wordq(1,4,3,2)\cr
&       + {1\over 12} \Wordq(2,1,3,4)
       - {1\over 12} \Wordq(2,1,4,3)
       + {1\over 12} \Wordq(2,3,1,4)
       + {1\over 12} \Wordq(2,3,4,1)
       + {1\over 12} \Wordq(2,4,1,3)
       - {1\over 12} \Wordq(2,4,3,1)\cr
&       + {1\over 12} \Wordq(3,1,2,4)
       - {1\over 12} \Wordq(3,1,4,2)
       - {1\over 12} \Wordq(3,2,1,4)
       - {1\over 12} \Wordq(3,2,4,1)
       + {1\over 12} \Wordq(3,4,1,2)
       - {1\over 12} \Wordq(3,4,2,1)\cr
&       + {1\over 12} \Wordq(4,1,2,3)
       - {1\over 12} \Wordq(4,1,3,2)
       - {1\over 12} \Wordq(4,2,1,3)
       - {1\over 12} \Wordq(4,2,3,1)
       - {1\over 12} \Wordq(4,3,1,2)
       - {1\over 4} \Wordq(4,3,2,1)
}$$

\newsubsecstar\BGexapp The Berends-Giele idempotents

The Berends-Giele idempotents $\cE(P)$ are defined in section~\BGeulersec\ 
as the inverse $\t(E(P))$ of the
Eulerian idempotent \eulsi. Their expansions up to multiplicity three were given in \cEexs\
and now we write down the multiplicity four:
\eqnn\cEfour
$$\eqalignno{
\cE(1234) &=
        {1\over4} \Wordq(1,2,3,4)
       - {1\over12} \Wordq(1,2,4,3)
       - {1\over12} \Wordq(1,3,2,4)
       - {1\over12} \Wordq(1,3,4,2)
       - {1\over12} \Wordq(1,4,2,3)
       + {1\over12} \Wordq(1,4,3,2) &\cEfour\cr
&       - {1\over12} \Wordq(2,1,3,4)
       + {1\over12} \Wordq(2,1,4,3)
       - {1\over12} \Wordq(2,3,1,4)
       - {1\over12} \Wordq(2,3,4,1)
       + {1\over12} \Wordq(2,4,1,3)
       + {1\over12} \Wordq(2,4,3,1)\cr
&       - {1\over12} \Wordq(3,1,2,4)
       - {1\over12} \Wordq(3,1,4,2)
       + {1\over12} \Wordq(3,2,1,4)
       + {1\over12} \Wordq(3,2,4,1)
       - {1\over12} \Wordq(3,4,1,2)
       + {1\over12} \Wordq(3,4,2,1)\cr
&       - {1\over12} \Wordq(4,1,2,3)
       + {1\over12} \Wordq(4,1,3,2)
       + {1\over12} \Wordq(4,2,1,3)
       + {1\over12} \Wordq(4,2,3,1)
       + {1\over12} \Wordq(4,3,1,2)
       - {1\over4} \Wordq(4,3,2,1)
}$$
As a curiosity, noting that $E_4=I_4$ one can derive these permutations using the conversion \IpToBp\
together with \BpToPermutations\ for the permutations in $\t(B_p)$ (note $\t^2=1$). So
\Ifour\ 
yields the permutations in $\cE_{1234}=\t(I_4)$ as
\eqnn\cEperms
$$\eqalignno{
\cE(1234)& = - {1\over4} 1\shuffle 2\shuffle3\shuffle4
       + {1\over3} 1\shuffle2\shuffle34
       + {1\over3} 1\shuffle 23\shuffle4
       - {1\over2} 1\shuffle 234 &\cEperms\cr&
       + {1\over3} 12\shuffle 3\shuffle 4
       - {1\over2} 12\shuffle 34
       - {1\over2} 123\shuffle4
       + 1234\,.
}$$

\appendix{D}{Parity of the disk amplitude and even partitions}
\applab\diskparityap

\noindent Parity of the amplitude $A^{\rm string}(1, \ldots,n) = (-1)^n A^{\rm string}(n, \ldots,1)$
explains the vanishing of $A_{\z_2^m\z_M}(\g_{1|P_1, \ldots,P_k})$ for even $k$ as observed in Table 1. A quick
counting argument suggests why this is so as
$\sum_{k }{n-1\stirling 2k} = \half(n-1)!$ is the upper bound in the dimension of string disk amplitudes
from properties of the string worldsheet alone \monodStieberger. More precisely:

\proclaim Proposition. If $k$ is even then the $n$-point disk amplitude satisfies
\eqn\propeven{
A^{\rm string}(\g_{1|P_1, \ldots,P_k}) = 0,
}
where $\g_{1|P_1, \ldots,P_k}$ is the BRST-invariant permutation \BRSTdef.\par
\noindent{\it Proof.}
The parity of $A^{\rm string}$ at $n$ points
can be written as $A^{\rm string}(1,\s) = (-1)^n A^{\rm string}(1, \tilde\s)$ by cyclicity.
This means, by \gammacIp,
that $A^{\rm string}(\g_{1|P_1, \ldots,P_k})$ will vanish whenever the parity of $A^{\rm string}$ at
$n$ points is opposite to the parity of $\cI_p$ for $p\models n{-}1$. To see why this is true consider
the example of $A^{\rm string}(\g_{1|23,4})$ with the expression for the BRST-invariant permutation
in \multfourgammas. The terms can
be rearranged as
\eqn\parityex{
A^{\rm string}(\g_{1|23,4}) =
	  {1\over2}\big( \Astring_{1234}
	  - \Astring_{1432}\big)
          + {1\over2}\big( \Astring_{1243}
          - \Astring_{1342}\big)
          + {1\over2}\big( \Astring_{1423}
	  - \Astring_{1324}\big) = 0
}
which vanishes by parity $\Astring_{1234}=\Astring_{1432}$. Notice that this happens because
the parity of $\cI_{21}(23,4)$ from \revcIp\ is the opposite of the string disk amplitude;
${\tilde \cI}_{21}(23,4) = -\cI_{21}(23,4)$.
The proposition can now be proven by considering the two cases when $n$ is even or odd.

For $n$ even the parity of the $n$-point disk amplitude is ${+}$ so
$A^{\rm string}(\g_{1|P_1, \ldots,P_k})$
vanishes if the parity of $\cI_{p}$ is ${-}$ for a composition $p$ of $n{-}1$.
By \revcIp\ this means that
there must be an odd
number of even parts in the composition $p$ (which sum to even). But since $n{-}1$ is odd, there must be an odd number of odd
parts in $p$ (which sum to odd). Therefore the number of parts $k(p)$ is even ($={\rm odd}+{\rm odd}$). Similarly, when $n$ is odd the number
of parts $k(p)$ in the composition of $p$ is also even (from ${\rm even} + {\rm even}$). This finishes the proof. \QED

\newskip\refskip\refskip13pt plus 1pt minus 1pt
\def\listrefs{\bigskip%
\immediate\closeout\rfile\writestoppt%
\baselineskip=\refskip\centerline{{\bf References}}\bigskip{\parindent=0pt%
\frenchspacing\escapechar=` \input \jobname.refs\vfill\eject}\nonfrenchspacing}

\ninerm
\listrefs
\bye